\pdfoutput=1

\documentclass[11pt,twoside,a4paper,cmspaper,final,collab]{cms-tdr}
\def\svnVersion{494222P}\def\cmsCernNoTag{CERN-EP-2018-290}\def\cmsCernDate{\today}\def\cmsMessage{Published in the European Physical Journal C as \href{http://dx.doi.org/10.1140/epjc/s10052-019-6855-8}{\doi{10.1140/epjc/s10052-019-6855-8.}}}

\begin{document}\cmsNoteHeader{B2G-17-012}

\hyphenation{had-ron-i-za-tion}
\hyphenation{cal-or-i-me-ter}
\hyphenation{de-vices}
\RCS$HeadURL: svn+ssh://svn.cern.ch/reps/tdr2/papers/B2G-17-012/trunk/B2G-17-012.tex $
\RCS$Id: B2G-17-012.tex 492178 2019-03-19 15:29:33Z alverson $

\ifthenelse{\boolean{cms@external}}{\providecommand{\cmsTable}[1]{#1}}{\providecommand{\cmsTable}[1]{\resizebox{\textwidth}{!}{#1}}}
\newlength\cmsTabSkip\setlength{\cmsTabSkip}{1ex}
\providecommand{\bW}{\ensuremath{\cPqb\PW}\xspace}
\providecommand{\tZ}{\ensuremath{\cPqt\cPZ}\xspace}
\providecommand{\tH}{\ensuremath{\cPqt\PH}\xspace}
\providecommand{\tW}{\ensuremath{\cPqt\PW}\xspace}
\providecommand{\bZ}{\ensuremath{\cPqb\cPZ}\xspace}
\providecommand{\bH}{\ensuremath{\cPqb\PH}\xspace}
\providecommand{\bbarW}{\ensuremath{\cPaqb\PW}\xspace}
\providecommand{\tbarZ}{\ensuremath{\cPaqt\cPZ}\xspace}
\providecommand{\tbarH}{\ensuremath{\cPaqt\PH}\xspace}
\providecommand{\tbarW}{\ensuremath{\cPaqt\PW}\xspace}
\providecommand{\bbarZ}{\ensuremath{\cPaqb\cPZ}\xspace}
\providecommand{\bbarH}{\ensuremath{\cPaqb\PH}\xspace}
\newcommand{\T}{\ensuremath{\cmsSymbolFace{T}}\xspace}
\newcommand{\B}{\ensuremath{\cmsSymbolFace{B}}\xspace}
\providecommand{\PV}{\ensuremath{\cmsSymbolFace{V}}\xspace}
\newcommand{\TTbar}{\ensuremath{\T}\ensuremath{\overline{\T}}\xspace}
\newcommand{\BBbar}{\ensuremath{\B}\ensuremath{\overline{\B}}\xspace}
\newcommand{\TtobW}{\ensuremath{\T \to \cPqb\PW}\xspace}
\newcommand{\TtotZ}{\ensuremath{\T \to \cPqt\cPZ}\xspace}
\newcommand{\TtotH}{\ensuremath{\T \to \cPqt\PH}\xspace}
\newcommand{\TtobWtZtH}{\ensuremath{\T \to \cPqb\PW, \cPqt\cPZ, \cPqt\PH}\xspace}
\newcommand{\BtotW}{\ensuremath{\B \to \cPqt\PW}\xspace}
\newcommand{\BtobZ}{\ensuremath{\B \to \cPqb\cPZ}\xspace}
\newcommand{\BtobH}{\ensuremath{\B \to \cPqb\PH}\xspace}
\newcommand{\BtotWbZbH}{\ensuremath{\B \to \cPqt\PW, \cPqb\cPZ, \cPqb\PH}\xspace}
\newcommand{\TTtotZtZ}{\ensuremath{\TTbar \to \cPqt\cPZ\cPqt\cPZ}\xspace}
\newcommand{\TTtotZtH}{\ensuremath{\TTbar \to \cPqt\cPZ\cPqt\PH}\xspace}
\newcommand{\TTtotZbW}{\ensuremath{\TTbar \to \cPqt\cPZ\cPqb\PW}\xspace}
\newcommand{\BBtobZbZ}{\ensuremath{\BBbar \to \cPqb\cPZ\cPqb\Z}\xspace}
\newcommand{\BBtobZbH}{\ensuremath{\BBbar \to \cPqb\cPZ\cPqb\PH}\xspace}
\newcommand{\BBtobZtW}{\ensuremath{\BBbar \to \cPqb\cPZ\cPqt\PW}\xspace}
\newcommand{\Zee}{\ensuremath{\PZ\to\EE}\xspace}
\newcommand{\Zmumu}{\ensuremath{\PZ\to\MM}\xspace}
\newcommand{\Zll}{\ensuremath{\PZ\to\ell^{+}\ell^{-}}\xspace}
\newcommand{\ee}{\ensuremath{\EE}\xspace}
\newcommand{\mumu}{\ensuremath{\MM}\xspace}
\newcommand{\qqpbar}{\ensuremath{\PQq\PAQq^\prime}\xspace}
\newcommand{\Wqq}{\ensuremath{\PW\to \qqpbar}\xspace}
\newcommand{\Zqq}{\ensuremath{\PZ\to \qqbar}\xspace}
\newcommand{\Vqq}{\ensuremath{\PV\to \qqbar}\xspace}
\newcommand{\Hbb}{\ensuremath{\PH\to \bbbar}\xspace}
\newcommand{\ttjets}{\ensuremath{\cPqt\overline{\cPqt}}+jets\xspace}
\newcommand{\Zjets}{\ensuremath{\PZ}+jets\xspace}
\newcommand{\ttZ}{\ensuremath{\cPqt\overline{\cPqt}\PZ}\xspace}
\newcommand{\ttW}{\ensuremath{\cPqt\overline{\cPqt}\PW}\xspace}
\newcommand{\tWZ}{\ensuremath{\cPqt\PW\PZ}\xspace}
\newcommand{\tZq}{\ensuremath{\cPqt\PZ\PQq}\xspace}
\newcommand{\GAMSTAR}{{\ensuremath{\gamma^{*}}}\xspace}
\newcommand{\GAMSTARJET}{{\ensuremath{\gamma^{*}}}+jets\xspace}
\newcommand{\bjet}{\cPqb~jet\xspace}
\newcommand{\bjets}{\cPqb~jets\xspace}
\newcommand{\ST}{\ensuremath{S_{\mathrm{T}}}\xspace}
\newcommand{\BR}{\mathcal{B}\xspace}
\newcommand{\BRs}{\mathcal{B}\mathrm{s}}
\newcommand{\BRSigma}{\mathcal{B} \sigma}
\newcommand{\intL}{35.9\fbinv}
\newcommand{\mtprime}{\ensuremath{M(\T)}\xspace}
\newcommand{\mbprime}{\ensuremath{M(\B)}\xspace}
\newcommand{\mpruned}{\ensuremath{m_\text{pruned}}\xspace}
\newcommand{\msd}{\ensuremath{m_\text{sd}}\xspace}
\newcommand{\nsubqq}{\ensuremath{\tau_{21}}\xspace}
\newcommand{\nsubqqq}{\ensuremath{\tau_{32}}\xspace}
\newcommand{\jone}{\ensuremath{j_{1}}\xspace}
\newcommand{\jtwo}{\ensuremath{j_{2}}\xspace}
\newcommand{\jthree}{\ensuremath{j_{3}}\xspace}

\cmsNoteHeader{B2G-17-012}

\title{Search for vector-like quarks in events with two oppositely charged leptons and jets in proton-proton collisions at \texorpdfstring{$\sqrt{s} = 13\TeV$}{sqrt(s) = 13 TeV}}
\titlerunning{Search for vector-like quarks in leptons and jets at $\sqrt{s}=13\TeV$}

\author*[ku]{Sadia Khalil}

\date{\today}

\abstract{
A search for the pair production of heavy vector-like partners \T and \B of the top and bottom quarks has been performed by the CMS experiment at the CERN LHC using proton-proton collisions at $\sqrt{s} = 13\TeV$. The data sample was collected in 2016 and corresponds to an integrated luminosity of 35.9\fbinv. Final states studied for \TTbar production include those where one of the \T quarks decays via \TtotZ and the other via \TtobW, \tZ, or \tH, where \PH is a Higgs boson. For the \BBbar case, final states include those where one of the \B quarks decays via \BtobZ and the other \BtotW, \bZ, or \bH. Events with two oppositely charged electrons or muons, consistent with coming from the decay of a \PZ boson, and jets are investigated. The number of observed events is consistent with standard model background estimations. Lower limits at 95\% confidence level are placed on the masses of the \T and \B quarks for a range of branching fractions. Assuming 100\% branching fractions for \TtotZ, and \BtobZ, \T and \B quark mass values below 1280 and 1130\GeV, respectively, are excluded.
}

\hypersetup{
pdfauthor={CMS Collaboration},
pdftitle={Search for vector-like quarks in events with two oppositely charged leptons and jets in proton-proton collisions at sqrt(s)=13 TeV},
pdfsubject={CMS},
pdfkeywords={CMS, vector-like quarks, pair-production, dileptons, 13 TeV}}

\maketitle

\section{Introduction\label{sec:Intro}}

The standard model (SM) has been outstandingly successful in describing a wide range of fundamental phenomena. However, one of its notable shortcomings is that it does not provide a natural explanation for the Higgs boson (\PH)~\cite{HiggsDiscoveryAtlas,HiggsDiscoveryCMS,CMSHiggsLongPaper} observed at 125\GeV~\cite{Aad:2015zhl,Sirunyan:2017exp} having a mass that is comparable to the electroweak scale. The suppression of divergent loop corrections to the Higgs boson mass requires either fine-tuning of the SM parameters or new particles at the \TeV scale. Many theories of beyond-the-SM physics phenomena that attempt to solve this hierarchy problem predict new particles, which could be partners of the top and bottom quarks and thus cancel the leading loop corrections. Vector-like quarks (VLQs) represent one class of such particles among those that have fermionic properties. Their left- and right-handed components transform in the same way under the SM symmetry group $\mathrm{SU(3)_C{\times}SU(2)_L{\times}U(1)_Y}$~\cite{VLQHandbookSaavedra}. This property allows them to have a gauge-invariant mass term in the Lagrangian of the form $\overline{\psi}\psi$, where $\psi$ represents the fermion field; hence, their masses are not determined by their Yukawa couplings to the Higgs boson. These quarks are not ruled out by the measured properties of the Higgs boson. They are predicted in many beyond-the-SM scenarios such as grand unified theories~\cite{GUTVLQ}, beautiful mirrors~\cite{beautiful}, models with extra dimensions~\cite{rs1}, little Higgs~\cite{littlesthiggs,Schmaltz200340,littlehiggsreview}, and composite Higgs models~\cite{compositehiggs}, as well as theories proposed to explain the SM flavor structure~\cite{gaugeflavor} and solve the strong CP problem~\cite{VLQstrongCPproblem}.

The VLQs can be produced singly or in pairs~\cite{VLQHandbookSaavedra}. The cross section for single-quark production is model dependent and depends on the couplings of the VLQs to the SM quarks. On the other hand, pair production of VLQs occurs via the strong interaction, and its cross section is uniquely determined by the mass of the VLQ. Another characteristic of the VLQs is their flavor-changing neutral current decay, which distinguishes them from chiral fermions. The top and bottom quark VLQ partners \T and \B are expected to couple to the SM third-generation quarks~\cite{Okada:2012gy}, and decay via \TtobWtZtH and \BtotWbZbH, respectively.

\begin{figure*}[hbtp]
\centering
    \includegraphics[width=0.49\textwidth]{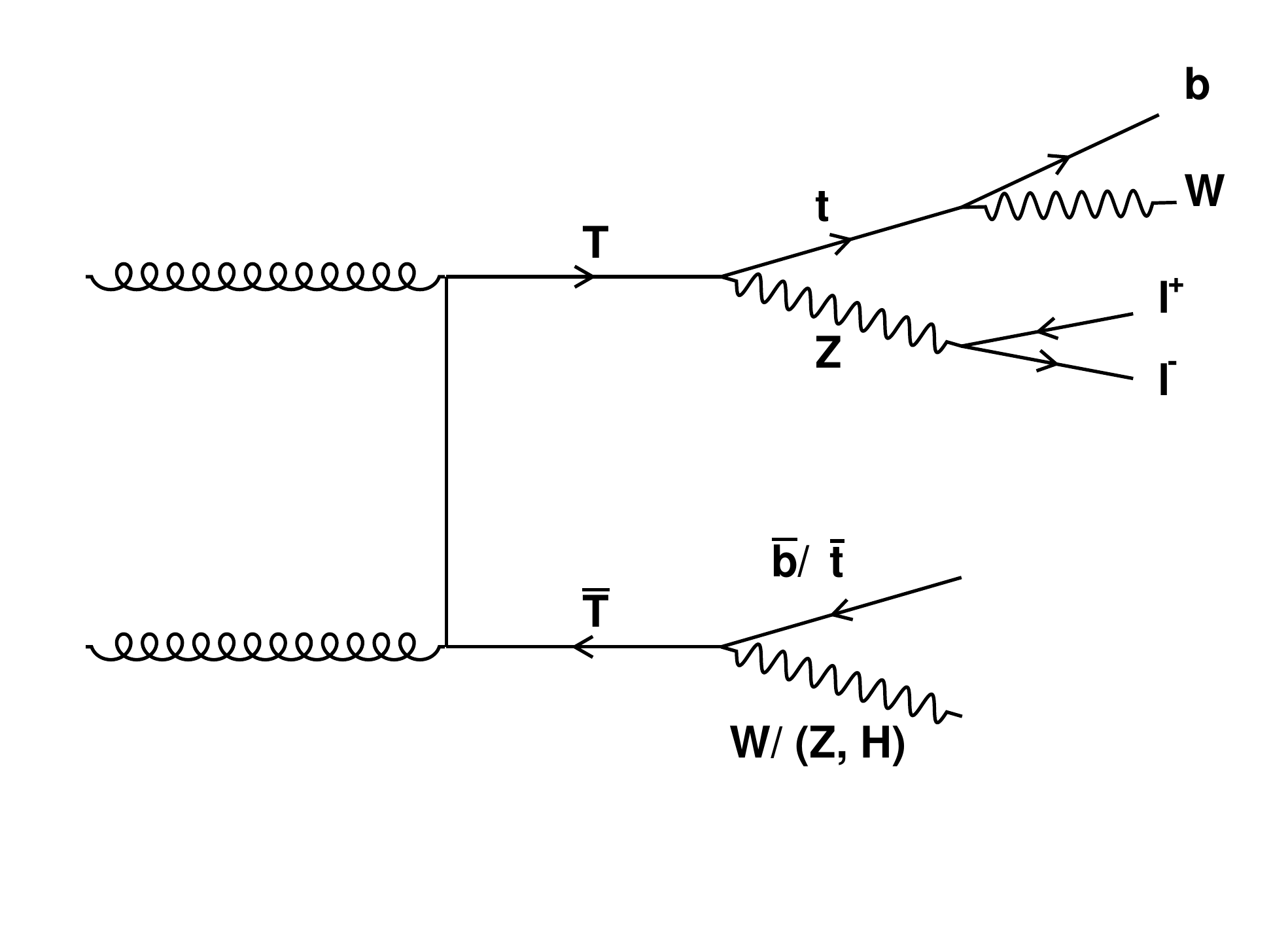}
    \includegraphics[width=0.49\textwidth]{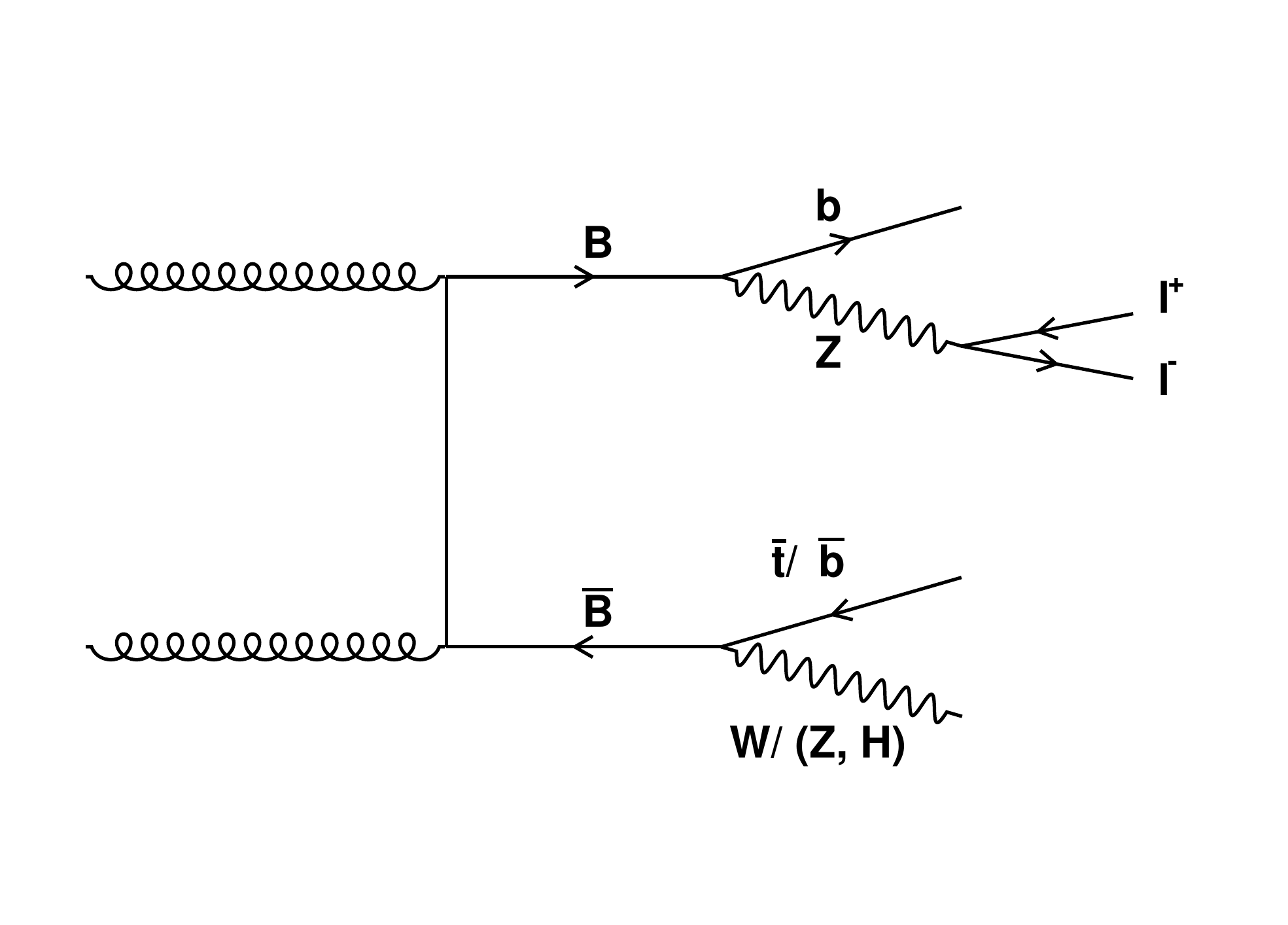}
    \caption{Leading-order Feynman diagrams for the pair production and decay of \T (left) and \B (right) VLQs relevant to final states considered in this analysis.}
    \label{fig:vlq}

\end{figure*}

In this paper, a search for the production of \TTbar and \BBbar is presented, where at least one of the \T (\B) quarks decays as \TtotZ (\BtobZ), as shown in Fig.~\ref{fig:vlq}. The search is performed using events with two oppositely charged electrons or muons, consistent with coming from a decay of a \PZ boson, and jets. The data were collected with the CMS detector at the CERN LHC in 2016, from proton-proton ({\Pp}{\Pp}) collisions at $\sqrt{s} = 13$\TeV, corresponding to an integrated luminosity of \intL.

Searches for the pair production of \T and \B quarks have previously been reported by the ATLAS~\cite{Aaboud:2017qpr,Aaboud:2017zfn,atlas_Jun7,ATLAS-new} and CMS~\cite{B2G-16-024,B2G-17-003,B2G-17-011} Collaborations. The strictest lower limits on the \T and \B quark masses range between 790 and 1350\GeV, depending on the decay mode studied. The mass range for the \T and \B quarks studied in this analysis is 800--1500\GeV.

\section{The CMS detector and event simulation\label{sec:CMS}}

The central feature of the CMS apparatus is a superconducting solenoid of 6\unit{m} internal diameter, providing a magnetic field of 3.8\unit{T}. Within the solenoid volume are a silicon pixel and strip tracker, a lead tungstate crystal electromagnetic calorimeter (ECAL), and a brass and scintillator hadron calorimeter (HCAL), each composed of a barrel and two endcap sections. Forward calorimeters extend the pseudorapidity ($\eta$) coverage provided by the barrel and endcap detectors. Muons are detected in gas-ionization chambers embedded in the steel flux-return yoke outside the solenoid. A more detailed description of the CMS detector, together with a definition of the coordinate system used and the relevant kinematic variables, can be found in Ref.~\cite{Chatrchyan:2008zzk}.

Events of interest are selected using a two-tiered trigger system~\cite{CMSTrigger}. The first level, composed of custom hardware processors, uses information from the calorimeters and muon detectors to select events at a rate of around 100\unit{kHz} within a time interval of less than 4\mus. The second level, known as the high-level trigger, consists of a farm of processors running a version of the full event reconstruction software optimized for fast processing, and reduces the event rate to around 1\unit{kHz} before data storage.

{\tolerance=1000
Monte Carlo (MC) simulated signal events of the processes ${\Pp}{\Pp}\to\TTbar$ and ${\Pp}{\Pp}\to\BBbar$ for \T and \B quark masses in the range 0.8--1.5\TeV are produced in steps of 0.1\TeV. The events are generated with \MGvATNLO 2.3.3~\cite{MG5_aMCNLO}, where the processes are produced at leading order (LO) with up to two partons in the matrix element calculations, using the {NNPDF3.0} parton distribution function (PDF) set~\cite{Ball:2014uwa}. Showering and hadronization is simulated with \PYTHIA~8.212~\cite{Pythia8p2} using the underlying event tune {CUETP8M1}~\cite{CUETTune}. To normalize the simulated signal samples to the data, next-to-next-to-leading-order (NNLO) and next-to-next-to-leading-logarithmic (NNLL) soft-gluon resummation cross sections are obtained using the \textsc{Top++} program (v.2.0)~\cite{Czakon:2013goa}, with the {MSTW2008NNLO68CL} PDF set as implemented in the {LHAPDF} (v.5.9.0) framework~\cite{LHAPDF}.
\par}

{\tolerance=800
The main background process is Drell--Yan (\cPZ/\GAMSTAR)+jets production, with smaller contributions from \ttjets and \ttZ. Throughout the paper this background will be referred to as DY+jets. Other backgrounds, such as diboson, \tZq, \tWZ, and \ttW production, are considerably smaller. The DY+jets simulated background samples are generated in different bins of the \PZ boson transverse momentum \pt, using the \MCATNLO\cite{MCATNLO} event generator at NLO precision with the \textsc{FxFx} jet-matching scheme~\cite{FXFX}. The \ttjets events are generated using the {\POWHEG~2.0}~\cite{POWHEG,POWHEG_Frixione,POWHEGBOX} generator. The generated events are interfaced with \PYTHIA~8.212~\cite{Pythia8p2} for shower modeling and hadronization, using the underlying event tune {CUETP8M2T4}~\cite{CMS-PAS-TOP-16-021} for \ttjets simulation and {CUETP8M1}~\cite{CUETTune} for the DY+jets process. The SM diboson events are also produced using the same standalone \PYTHIA~8.212 generator. The production of rare single top processes \tZq and \tWZ, as well as a \ttbar pair in association with a \PW or \PZ boson, are simulated with up to one additional parton in the matrix element calculations using the \MGvATNLO 2.3.3~\cite{MG5_aMCNLO} generator at LO precision and matched with the parton showering predictions using the MLM matching scheme~\cite{MLM}.
\par}

{\tolerance=800
Backgrounds are normalized according to the theoretical predictions for the corresponding cross sections. The DY+jets production cross sections from the \MCATNLO~\cite{MCATNLO} generator are valid up to NLO. Using a top quark mass of 172.5\GeV, the \ttjets production cross section at NNLO~\cite{Czakon:2013goa} is determined. Diboson production is calculated at NLO for $\PW\PZ$~\cite{WZ} and NNLO for $\PZ\PZ$~\cite{ZZ} and $\PW\PW$~\cite{WW}. The production cross sections for the rare processes \tZq, \tWZ, and \ttW are calculated at NLO~\cite{LHCHXSecWG}.
\par}

{\tolerance=800
A \GEANTfour-based~\cite{Agostinelli:2002hh,GEANT} simulation of the CMS apparatus is used to model the detector response, followed by event reconstruction using the same software configuration as for the collision data. The effect of additional {\Pp}{\Pp} interactions in the same or nearby bunch crossings (pileup) in concurrence with the hard scattering interaction is simulated using the \PYTHIA~8.1 generator and a total inelastic {\Pp}{\Pp} cross section of 69.2\unit{mb}~\cite{LHCHXSecWG}. The frequency distribution of the additional events is adjusted to match that observed in data and has a mean of 23.
\par}

\section{Event reconstruction\label{sec:EvtReco}}

{\tolerance=800
The event reconstruction in CMS uses a particle-flow (PF) algorithm~\cite{CMS-PRF-14-001} to reconstruct a set of physics objects (charged and neutral hadrons, electrons, muons, and photons) using an optimized combination of information from the subdetectors. The energy calibration is performed separately for each particle type.
\par}

{\tolerance=800
The {\Pp}{\Pp} interaction vertices are reconstructed from tracks in the silicon tracker using the deterministic annealing filter algorithm~\cite{PV}. The {\Pp}{\Pp} interaction vertex with the highest $\sum\pt^{2}$ of the associated clusters of physics objects is considered to be the primary vertex associated with the hard scattering interaction. Here, the physics objects are the jets, which are clustered with the tracks assigned to the vertex using the anti-\kt jet clustering algorithm~\cite{antiKtAlgorithm,FastJet}, and the missing transverse momentum \ptvecmiss, defined as the negative vector sum of the \ptvec of those jets, with its magnitude referred to as \ptmiss. The interaction vertices not associated with the hard scattering are designated as pileup vertices.\par}

Electron candidates are reconstructed from clusters of energy deposited in the ECAL and from hits in the silicon tracker~\cite{Electrons}. The clusters are first matched to track seeds in the pixel detector, then the trajectory of an electron candidate is reconstructed considering energy lost by the electron due to bremsstrahlung as it traverses the material of the tracker, using a Gaussian sum filter algorithm. The PF algorithm further distinguishes electrons from charged pions using a multivariate approach~\cite{PFT-10-001}. Observables related to the energy and geometrical matching between track and ECAL cluster(s) are used as main inputs. Additional requirements are applied on the ECAL shower shape, the variables related to the track-cluster matching, the impact parameter, and the ratio of the energies measured in the HCAL and ECAL in the region around the electron candidate. With these requirements, the reconstruction and identification efficiency of an electron from a \Zee decay is on average 70\%, whereas the misidentification rate is 1--2\%~\cite{Electrons}. Electrons with $\pt > 25\GeV$ and $\abs{\eta}<2.4$ are selected for this analysis. Further, electrons passing through the transition regions between the ECAL barrel and endcap sections, ($1.444 < \abs{\eta} < 1.566$), which are less well measured, are removed.

Muon candidates are identified by multiple reconstruction algorithms using hits in the silicon tracker and signals in the muon system. The standalone muon algorithm uses only information from the muon detectors. The tracker muon algorithm starts from tracks found in the silicon tracker and then associates them with matching tracks in the muon detectors. The global muon algorithm starts from standalone muons and then performs a global fit to consistent hits in the tracker and the muon system~\cite{Sirunyan:2018}. Global muons are used by the PF algorithm. Muons are required to pass additional identification criteria based on the track impact parameter, the quality of the track reconstruction, and the number of hits recorded in the tracker and the muon systems. Muons selected for this analysis are required to have $\pt > 25\GeV$ and $\abs{\eta}<2.4$.

Charged leptons (electrons or muons) from \Zee or \Zmumu decays, with the \PZ boson originating from the decay of a heavy VLQ, are expected to be isolated, \ie, to have low levels of energy deposited in the calorimeter regions around their trajectories. An isolation variable is defined as the scalar \pt sum of the charged and neutral hadrons and photons in a cone centered on the direction of the lepton, of radius $\Delta R \equiv \sqrt{\smash[b]{(\Delta \eta)^2+(\Delta \phi)^2}}$, with $\Delta R = 0.3\,(0.4)$ for electrons (muons). The \pt contributions from pileup and from the lepton itself are subtracted from the isolation variable~\cite{Electrons,Sirunyan:2018}. The relative isolation parameter, defined as the isolation variable divided by the lepton \pt, is required to be less than 0.06 (0.15) for the electrons (muons), with corresponding efficiencies of 85 and 95\%, respectively, based on simulation. The isolation requirement helps reject jets misidentified as leptons and reduce multijet backgrounds.

The anti-\kt jet clustering algorithm~\cite{antiKtAlgorithm,FastJet} reconstructs jets with PF candidates as inputs. The energy of charged hadrons is determined from a combination of their momentum measured in the tracker and the matching ECAL and HCAL energy deposits, corrected for zero-suppression effects and for the response function of the calorimeters to hadronic showers. Finally, the energy of neutral hadrons is obtained from the corresponding corrected ECAL and HCAL energies. To suppress the contribution from pileup, charged particles not originating from the primary vertex are removed from the jet clustering. An event-by-event jet-area-based correction~\cite{PUSubJetArea,CMSJECJER} is applied to subtract the contribution of the neutral-particle component of the pileup. Residual corrections are applied to the data to account for the differences with the simulations~\cite{Khachatryan:2016kdb}.

Two types of jet are considered, distinguished by the choice of distance parameter used for clustering. Those clustered with a distance parameter of 0.4 (``AK4 jets"), are required to have $\pt > 30\GeV$, and those clustered with a value of 0.8 for this parameter (``AK8 jets") must satisfy the condition $\pt > 200\GeV$, where the jet momentum is the vector sum of the momenta of all particles clustered in the jet. Both classes of jets must satisfy $\abs{\eta}<2.4$. A new value for \ptmiss is determined using the PF objects and including the jet energy corrections.

The combined secondary vertex \cPqb\ tagging algorithm (CSVv2)~\cite{BTV-16-002} is used to identify jets originating from the hadronization of \cPqb\ quarks. The algorithm combines information on tracks from the silicon tracker and vertices associated with the jets using a multivariate discriminant. An AK4 jet is defined as a \cPqb-tagged jet if the corresponding CSVv2 discriminant is above a threshold that gives an average efficiency of about 70\% for \cPqb\ quark jets and a misidentification rate of 1\% for light-flavored jets.

The signal events searched for in this analysis have two massive VLQs decaying to at least one \PZ boson and either a \PZ, \PW, or Higgs boson and two heavy quarks. One \PZ boson must decay leptonically, whereas the remaining \PZ, \PW, or Higgs boson is reconstructed using its hadronic decays into jets. Depending on the mass of the VLQ, the decay products can have a large Lorentz boost. In this case, the decay products of \Wqq and \Zqq (collectively labeled as \Vqq), \Hbb, and $\cPqt\to\qqpbar\cPqb$ may be contained within a single AK8 jet. These decays are reconstructed using a jet substructure tagger. The decay products of heavy bosons and top quarks that do not acquire a large Lorentz boost are identified by a resolved tagger using AK4 jets. Both types of taggers are described in the next section.

\section{Event selection and categorization\label{sec:EvtSel}}

For the dielectron (\Zee) channel, event candidates are selected using triggers requiring the presence of at least one electron with $\pt>115\GeV$ or a photon with $\pt>175\GeV$. After passing one of the triggers, the triggering electron is also required to pass a set of criteria based on the electromagnetic shower shape and the quality of the electron track. A loose isolation criterion on the electrons is further required, as described in Section~\ref{sec:EvtReco}. One of the electrons is required to have $\pt>120\GeV$ in order to remain above the triggering electron \pt threshold. Since the signal electrons originate from the decay of highly boosted \PZ bosons, these selection criteria preserve the high signal efficiency, while reducing the number of misidentified electrons. The photon trigger helps to retain electrons with $\pt>300\GeV$ that would otherwise be lost because of the requirements on electromagnetic shower shape in the ECAL.

For the dimuon (\Zmumu) channel, event candidates are selected using a trigger that requires presence of at least one muon with $\pt>24\GeV$. The trigger implements a loose isolation requirement by allowing only a small energy deposit in the calorimeters around the muon trajectory. After passing the trigger, one of the muons from the \Zmumu decay must have $\pt>45\GeV$, which provides the largest background rejection that can be obtained without decreasing the signal efficiency for the VLQ mass range of interest. The trigger and lepton reconstruction and identification efficiencies are determined using a tag-and-probe method~\cite{CMSMuonReco}. Scale factors are applied to the simulated events to account for any efficiency differences between the data and simulation.

The invariant mass of the lepton pair from the \PZ boson leptonic decay must satisfy $75 < m(\ell\ell) < 105\GeV$, to be consistent with the \PZ boson mass, and have a total $\pt(\ell\ell) > 100\GeV$, appropriate for the decay of a massive VLQ. Events must have exactly one $\ee$ or $\mumu$ pair candidate consistent with a \PZ boson decay.

Events are required to have at least three AK4 jets with $\HT > 200\GeV$, and $\HT \equiv \sum\pt$, where the summation is over all jets in the event. The highest \pt (leading) AK4 jet is required to have $\pt > 100\GeV$, the second-highest-\pt (subleading) AK4 jet to have $\pt > 50\GeV$, and all other jets must satisfy the condition $\pt > 30\GeV$. The AK4 (AK8) jets j within $\Delta R(\ell, j) < 0.4$ $(0.8)$ of either lepton from the \PZ boson decay are not considered further in the analysis. At least one \cPqb-tagged jet with $\pt > 50\GeV$ is required. The \ST variable, defined as the sum of \HT, $\pt(\PZ)$, and \ptmiss, must be greater than 1000\GeV. The selection criteria are summarized in Table~\ref{tab:sel}. The selections are optimized to obtain the largest suppression of SM backgrounds that can be achieved without reducing the simulated signal efficiency by more than 1\%.

\begin{table}[!hbtp]
  \centering
    \topcaption{Event selection criteria.}
    \begin{tabular}{lr}
    \hline
      Variable & Selection \\
      \hline
      $\PZ\to\ell\ell$ candidate multiplicity & $=$1 \\
      $\pt(\PZ)$                              & $>$100\GeV \\
      AK4 jet multiplicity                    & $\ge$3 \\
      \HT                                     & $>$200\GeV \\
      \pt of leading AK4 jet                  & $>$100\GeV \\
      \pt of subleading AK4 jet               & $>$50\GeV \\
      \cPqb-tagged AK4 jet multiplicity       & $\ge$1 \\
      \pt of \cPqb\ jet                       & $>$50\GeV \\
      \ST                                     & $>$1000\GeV \\
   \hline
   \end{tabular}
   \label{tab:sel}

\end{table}

The event topologies are different for \TTbar and \BBbar decays, and the product of the signal efficiency and the acceptance varies from 1.2 to 2.6\% over the various signal channels. The \TTbar events are characterized by three heavy bosons and two heavy quarks in the decay sequence. The \BBbar events have two heavy bosons and two heavy quarks, hence more energetic final decay objects. Therefore, the analysis is optimized separately for the \TTbar and \BBbar channels.

For both searches the decays of boosted \Vqq and \Hbb are reconstructed from AK8 jets, using the jet substructure tagger, and are referred to as \PV and \PH jets, respectively. As the Higgs boson mass is larger than \PW\ and \PZ boson masses, it requires a higher momentum for its decay products to merge into a single AK8 jet. Therefore, \PH jets are required to have $\pt > 300\GeV$ and \PV jets have $\pt > 200\GeV$. A jet pruning algorithm~\cite{pruning,pruning1} is used to measure the jet mass. The \PV and \PH jet candidates are required to have a pruned jet mass in the range 65--105 and 105--135\GeV, respectively. The jet pruning algorithm reclusters the groomed jets~\cite{Salam:2009jx} by eliminating low energy subjets subjets. In the subsequent recombination of two subjets, the ratio of the subleading subjet \pt to the pruned jet \pt must be greater than 0.1 and the distance between the two subjets must satisfy $\Delta R < m_\mathrm{jet}/2{\pt}_\mathrm{jet}$, where $m_\mathrm{jet}$ and ${\pt}_\mathrm{jet}$ are the mass and \pt of the pruned jet, respectively.

The $N$-subjettiness algorithm~\cite{Thaler:2010tr} is used to calculate the jet shape variable $\tau_{N}$, which quantifies the consistency of a jet with the hypothesis of the jet having $N$ subjets, each arising from a hard parton coming from the decay of an original heavy boson. The \PV and \PH jets in the \TTbar (\BBbar) search are required to have an $N$-subjettiness ratio $\nsubqq \equiv \tau_{2}/\tau_{1} < 1.0\,(0.6)$. Both pruned subjets coming from the \PH jet are required to be \cPqb-tagged. This is done by using the above-mentioned CSVv2 b-tagging algorithm with a cut that gives a 70--90\% efficiency for \cPqb\ quark subjets, depending on the subjet \pt, and a misidentification rate of 10\% for subjets from light-flavored quarks and gluons.

Boosted top quarks decaying to $\cPqb\qqpbar$ are identified (``{\cPqt} tagged") using AK8 jets and the soft-drop algorithm~\cite{Dasgupta:2013ihk,softdrop} to groom the jet. This algorithm recursively declusters a jet into two subjets. It discards soft and wide-angle radiative jet components until a hard-splitting criterion is met, to obtain jets consistent with the decay of a massive particle. We use the algorithm with an angular exponent $\beta = 0$, a soft cutoff threshold $z_{cut} < 0.1$, and a characteristic radius $R_0 = 0.8$. For top quark jets, the soft-drop mass must be in the range 105--220\GeV and the $N$-subjettiness ratio $\nsubqqq \equiv \tau_{3}/\tau_{2} < 0.81\,(0.67)$ for the \TTbar (\BBbar) search, consistent with the expectation for three subjets from top quark decay. There are a total of five heavy bosons and quarks produced in \TTbar signal events, whereas there are only four in \BBbar events. Thus it is possible to apply a tighter $N$-subjetiness ratio criterion in the \BBbar analysis without a loss of signal efficiency.

Corrections to the jet mass scale, resolution and $\nsubqq$ selection efficiency for \PV jets due to the difference in data and MC simulation are measured using a sample of semileptonic \ttbar events~\cite{JME-16-003}. For the correction to the jet mass scale and resolution, boosted $\PW$ bosons produced in the top quark decays are separated from the combinatorial \ttbar background by performing a simultaneous fit to the observed pruned jet mass spectrum. In order to account for the difference in the jet shower profile of \Vqq and \Hbb decays, a correction factor to the \PH jets mass scale and resolution~\cite{H-taggingUnc} is measured by comparing the ratio of $\PH$ and $\PV$ jet efficiencies using the \PYTHIA~8.212~\cite{Pythia8p2} and \HERWIG{++}~\cite{Herwigpp} shower generators. In addition, the corrections to $\nsubqq$ selection efficiency are obtained based on the difference between data and simulation~\cite{H-taggingUnc} for \PH-tagged jets. All these corrections are propagated to \PV, top quark and \PH jets, respectively. For top quark jets, the corrections to the $\nsubqqq$ selection efficiency are measured between data and simulation~\cite{JME-16-003} using soft-drop groomed jets. To account for the misidentification of boosted \PV-, \PH-, and \cPqt-tagged jets in the background samples, mistagging scale factors are derived from a region in the data enriched in \Zjets events, which is constructed using the selection criteria listed in Table~\ref{tab:sel}, with the exception that events must have zero \cPqb\ jets. These mistagging scale factors are applied to the mistagged jets in simulated signal and background events.

In the \TTbar search, in addition to the jet substructure techniques, the \PW, \PZ, \PH, and top quark decays are reconstructed with a resolved tagger using AK4 jets, as described below. Only those AK4 jets that are a radial distance $\Delta R > 0.8$ from the tagged AK8 jets are considered in the resolved tagging algorithm. The resolved \Vqq and \Hbb candidates are composed of two AK4 jets \jone and \jtwo whose invariant mass must satisfy $70 < m(\jone\jtwo) < 120\GeV$ and $80 < m(\jone\jtwo) < 160\GeV$, respectively, and have $\pt(\jone\jtwo) > 100\GeV$. For \PH candidates, at least one of the jets must be \cPqb\ tagged. The resolved top quark candidate is composed of either three AK4 jets \jone, \jtwo, and \jthree with an invariant mass $120 < m(\jone\jtwo\jthree) < 240\GeV$ and $\pt(\jone\jtwo\jthree) > 100\GeV$, or an AK4 jet \jone and an AK8 \PV jet satisfying $120 < m(\PV\jone) < 240\GeV$ and $\pt(\PV\jone) > 150\GeV$. These selection criteria are derived from simulated \TTbar events, using MC truth information.

The \TTbar events are next classified based on the number of AK4 \cPqb-tagged jets ($N_{\cPqb}$), and number of \Vqq ($N_{\PV}$), \Hbb ($N_{\PH}$), and $\cPqt\to\qqpbar\cPqb$ ($N_{\cPqt}$) candidates identified using either the jet substructure or resolved tagging algorithms. In an event, $N_{\cPqb}$ can be 1 or ${\geq}2$, and $N_{\PV}$, $N_{\PH}$, and $N_{\cPqt}$ each can be 0 or ${\geq}1$. Thus, in total, $2{\times}2{\times}2{\times}2=16$ categories of events are constructed. For simplicity, overlaps between candidates of different types are allowed, \eg, the same AK8 jet could be tagged as both a top quark and an \PH candidate because of the overlapping mass windows. Such overlaps occur in a few percent of the signal events. However, by construction each event can belong to only one category. In the example above, the event would fall into a category with both $N_{\PH} \geq 1$ and $N_{\cPqt} \geq 1$ requirements satisfied. Further, the mistag rates and the relevant corrections to the jet mass scale and resolution are applied to the \PH and \cPqt\ candidates, based on MC truth information.

\begin{table*}[!htbp]
  \centering
    \topcaption{The first four columns show different event groups used for the \TTbar search, classified according to the number of \cPqb-tagged jets $N_{\cPqb}$ and the number of \Vqq, \Hbb, and $\cPqt\to\qqpbar\cPqb$ candidates in the event, $N_{\PV}$, $N_{\PH}$ and $N_{\cPqt}$, respectively, identified using both the jet substructure and resolved tagger algorithms. The last three columns show the relative signal acceptance for a \T quark of mass 1200\GeV for decay channels tZtZ, tZtH and tZbW as described in text.}
  \label{tab:TTevtgrp}
  \begin{tabular}{l r r r r r r r}
  \hline
    Group              & $N_{\cPqb}$ & $N_{\PV}$  & $N_{\PH}$ & $N_{\cPqt}$ & tZtZ (\%) & tZtH (\%) & tZbW (\%) \\
    \hline
    \multirow{2}{*}{A} & $=$1   & $\ge$1 & $=$0   & $\ge$1  & \multirow{2}{*}{37.8} & \multirow{2}{*}{27.2} & \multirow{2}{*}{31.9} \\
                       & $=$1   & $\ge$1 & $\ge$1 & $\ge$1  &                       &                       &  \\[\cmsTabSkip]
    \multirow{2}{*}{B} & $\ge$2 & $\ge$1 & $=$0   & $\ge$1  & \multirow{2}{*}{32.2} & \multirow{2}{*}{42.1} & \multirow{2}{*}{20.6} \\
                       & $\ge$2 & $\ge$1 & $\ge$1 & $\ge$1  &                       &                       &  \\[\cmsTabSkip]
    \multirow{2}{*}{C} & $=$1   & $=$0   & $=$0   & $\ge$1  & \multirow{2}{*}{8.4}  & \multirow{2}{*}{6.6}  & \multirow{2}{*}{11.6} \\
                       & $\ge$2 & $\ge$1 & $=$0   & $=$0    &                       &                       &  \\[\cmsTabSkip]
    \multirow{3}{*}{D} & $\ge$2 & $\ge$1 & $\ge$1 & $=$0    & \multirow{3}{*}{8.7}  & \multirow{3}{*}{13.4} & \multirow{3}{*}{8.2} \\
                       & $\ge$2 & $=$0   & $\ge$1 & $\ge$1  &                       &                       &  \\
                       & $\ge$2 & $=$0   & $=$0   & $\ge$1  &                       &                       &  \\
  \hline
  \end{tabular}

\end{table*}

Next, the event categories are sorted using the figure of merit $S/\sqrt{B}$, where $S$ and $B$ are the expected \TTtotZtZ signal and background event yields, respectively, as determined from the simulation. The categories with similar figures of merit based on expected upper limits at 95\% confidence level (\CL) are grouped together, while the categories that are found not to add sensitivity to the \TTbar signal are discarded. A total of four event groups labeled A through D are selected, each with a different signal acceptance relative to the selection criteria described in Table~\ref{tab:sel} and depending on the \T decay channel. Table~\ref{tab:TTevtgrp} shows the selections on these event groups, and the relative signal acceptances of the \T quark decay channels, namely, tZtZ, tZtH, or tZbW for a \T quark of mass 1200\GeV. The decay channels are defined with a benchmark combination of branching fractions $\BR(\TtotZ) = 100\%$ (tZtZ), $\BR(\TtotZ) = \BR(\TtotH) = 50\%$ (tZtH), and $\BR(\TtotZ) = \BR(\TtobW) = 50\%$ (tZbW). Events from all the decay channels mainly contribute to groups A and B, whereas groups C and D have slightly lower acceptance depending on the decay channel. The fraction of the signal identified by the jet substructure and resolved taggers depends on the \T quark mass. For masses below 1200\GeV, the two taggers are equally efficient in identifying signal events for all the channels. For \T quark masses above 1200\GeV, the jet substructure tagger becomes more efficient. For example, for \T quark mass at 1800\GeV, the jet substructure tagger selects twice as many \T quark candidates as the resolved tagger.

\begin{table*}[!htbp]
  \centering
  \topcaption{The first four columns show different event categories used for the \BBbar search, classified according to the number of AK4 \cPqb-tagged jets $N_{\cPqb}$ and the number of \Vqq, \Hbb, and $\cPqt\to\qqpbar\cPqb$ candidates in the event, $N_{\PV}$, $N_{\PH}$, and $N_{\cPqt}$, respectively, identified using the jet substructure algorithm. The last three columns show the relative signal acceptance for a \B quark of mass 1200\GeV for decay channels bZbZ, bZbH and bZtW as described in text.}
  \label{tab:BBevtgrp}
  \begin{tabular}{l r r r r r r r}
  \hline
    Category	  & $N_{\cPqb}$ & $N_{\PV}$   & $N_{\PH}$   & $N_{\cPqt}$ & bZbZ (\%) & bZbH (\%) & bZtW (\%)\\
    \hline
    1\cPqb\       & $=$1       & $=$0       & $=$0       & $=$0    & 50.4  & 27.4 &  22.3\\
    2\cPqb\       & $\ge$2     & $=$0       & $=$0       & $=$0    & 45.7  & 34.3 &  20.0\\
    Boosted \cPqt & $\ge$1     & $\ge$0     & $\ge$0     & $\ge$1  & 35.1  & 24.3 &  40.6\\
    Boosted \PH   & $\ge$1     & $\ge$0     & $\ge$0     & $=$0    & 21.4  & 64.3 &  14.3\\
    Boosted \PZ   & $\ge$1     & $\ge$1     & $=$0       & $=$0    & 52.4  & 21.7 &  25.9\\
  \hline
  \end{tabular}
  
\end{table*}

Because the event topology of \BBbar signal events is different from that of \TTbar signal events, as discussed previously, the \PV, \PH, and \cPqt\ candidates in the \BBbar analysis are identified using only the jet substructure tagger. Events are then separated into five categories, labeled 1\cPqb, 2\cPqb, boosted \cPqt, boosted \PH, and boosted \PZ, based on the values of $N_{\cPqb}$, $N_{\PV}$, $N_{\PH}$, and $N_{\cPqt}$. Table~\ref{tab:BBevtgrp} shows these categories, and the relative signal acceptances of \B quark decay channels, namely, bZbZ, bZbH, or bZtW for a \B quark of mass 1200\GeV. The decay channels are defined with a benchmark combination of branching fractions $\BR(\BtobZ) = 100\%$ (bZbZ), $\BR(\BtobZ) = \BR(\BtobH) = 50\%$ (bZbH), and $\BR(\BtobZ) = \BR(\BtotW) = 50\%$ (bZtW).

\section{Background modeling\label{sec:Bkg}}

The backgrounds from all sources are estimated using simulation, except for \Zjets where corrections to the simulated events are applied using data, as described below. The modeling of simulated background events is validated using several control regions in the data, which are constructed by inverting one or more of the requirements listed in Table~\ref{tab:sel}. The control region labeled CR0b+high-\ST is constructed by requiring zero \cPqb\ jets. The control region CR1\cPqb+low-\ST is constructed by inverting the \ST requirement: $\ST \le 1000\GeV$. The control region CR0b is constructed by requiring zero \cPqb\ jets and removing the \ST requirement. Signal contamination from all channels in each of these control regions is less than 1\%.

\begin{figure*}[!htbp]
\centering
\includegraphics[width=0.49\textwidth]{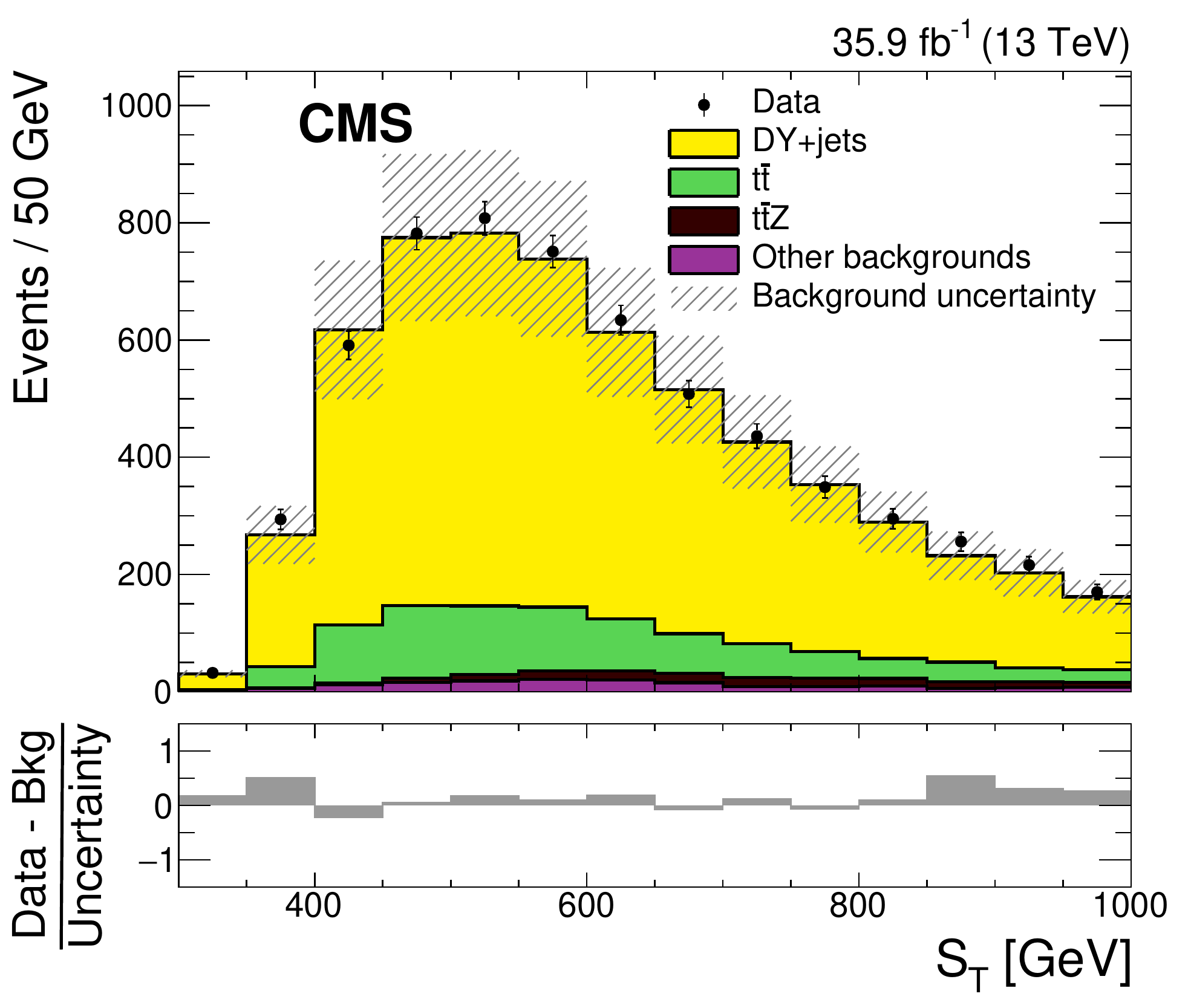}
\includegraphics[width=0.49\textwidth]{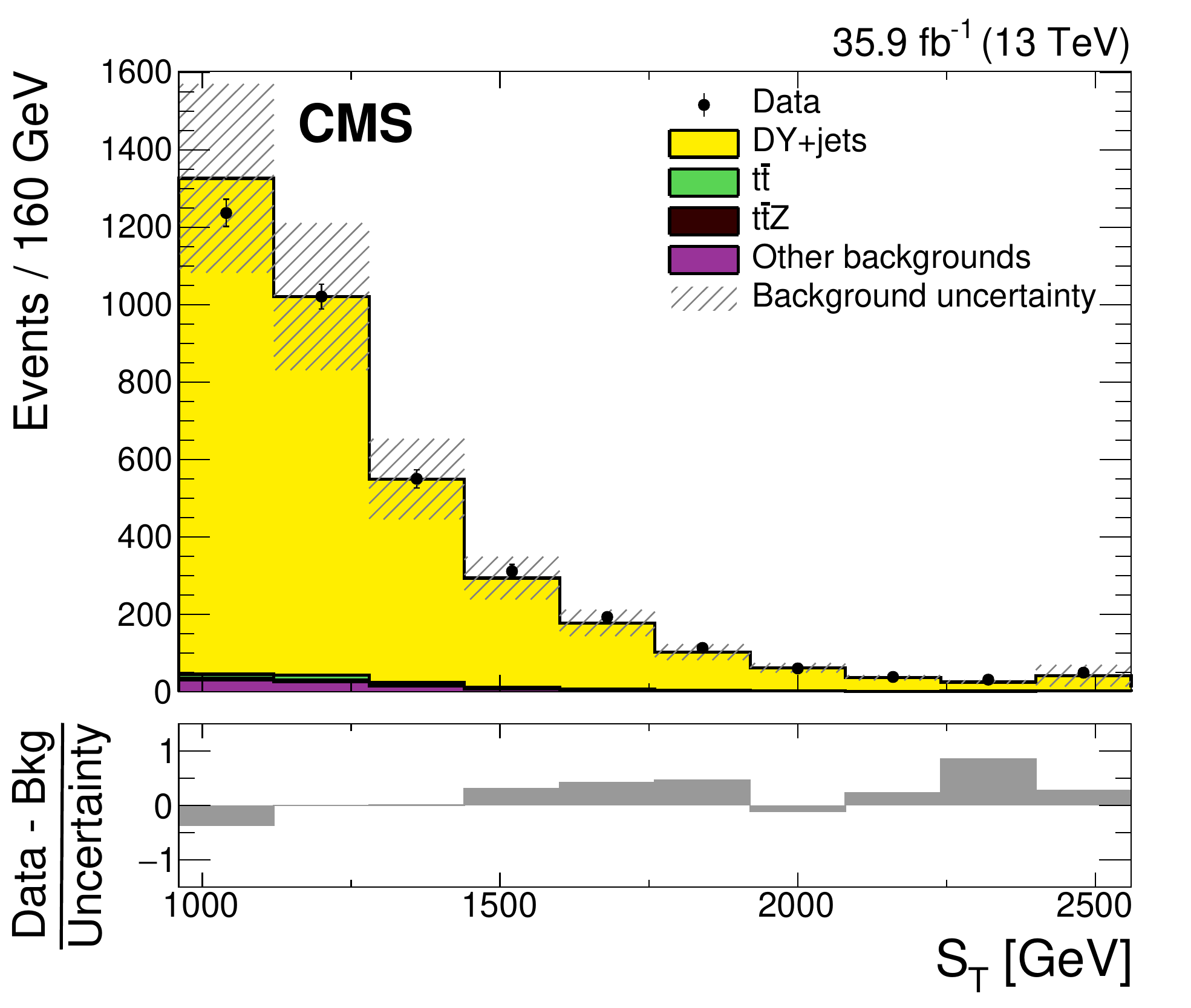}
\caption{The \ST distributions for the CR1\cPqb+low-\ST (left) and CR0\cPqb+high-\ST (right) control regions for the data (points) and the background simulations (shaded histograms) after applying the scale factors given in Table~\ref{tab:nak4w}. The vertical bars on the points represent the statistical uncertainties in the data. The hatched bands indicate the total uncertainties in the simulated background contributions added in quadrature. The lower plots show the difference between the data and the simulated background, divided by the total uncertainty.}
\label{fig:cnt_st}

\end{figure*}

The AK4 jet multiplicity distribution is not modeled reliably in the \Zjets simulation, and therefore it is corrected using scale factors obtained from data. Scale factors listed in Table~\ref{tab:nak4w} are determined using the CR0b control region, which is enriched with \Zjets events. After applying these corrections, the distributions of kinematic variables in the control regions from the background simulations are in agreement with the data, as shown for example in Fig.~\ref{fig:cnt_st} for the \ST distributions.

\begin{table}[!htbp]
  \centering
    \topcaption{The scale factors determined from data for correcting the AK4 jet multiplicity distribution in the simulation. The quoted uncertainties in the scale factors are statistical only.}
    \label{tab:nak4w}
    \begin{tabular}{l r r }
      \hline
      Number of AK4 jets  &  Scale factor    \\
      \hline
      3                  & 0.92 $\pm$ 0.01 \\
      4                  & 1.03 $\pm$ 0.01 \\
      5                  & 1.12 $\pm$ 0.02 \\
      6                  & 1.30 $\pm$ 0.05 \\
     ${\geq}7$           & 1.61 $\pm$ 0.12 \\
      \hline
    \end{tabular}

\end{table}

\section{Systematic uncertainties\label{sec:Syst}}

\begin{table*}[!htb]
\centering
    \topcaption{Summary of systematic uncertainties considered in the statistical analysis of \TTbar and \BBbar search on the background and signal events. All uncertainties affect the normalizations of the \ST distributions. The tick mark indicates the uncertainties that also affect the shape, and the uncertainty range accounts for their effects on the expected yields across all the \TTbar groups or \BBbar categories. The \TTbar and \BBbar signal events correspond to the benchmark decay channels tZtZ and bZbZ, respectively, for \T and \B quark mass $m_{\T}=m_{\B}=1200\GeV$.
}
    \label{tab:sys}
    \cmsTable{\begin{tabular}{c c c c c c }
      \hline
       \multicolumn{1}{c}{Source}& Shape & \multicolumn{4}{c}{Uncertainty (\%)}\\
       \hline
                                 &  & \multicolumn{2}{c}{\TTbar} & \multicolumn{2}{c}{\BBbar} \\
                                 &  &  Background yield & Signal yield & Background yield & Signal yield   \\
       $\ttbar$+jets rate        &  & 15                & ---           & 15               & ---              \\
       DY+jets rate              &  & 15                & ---           & 15               & ---              \\
       Diboson rate              &  & 15                & ---           & 15               & ---              \\
       Integrated luminosity     &  & 2.5               & 2.5          & 2.5              & 2.5             \\
       Lepton identification     &  & 3                 & 3            & 3                & 3               \\
       Trigger efficiency        &  & 1                 & 1            & 1                & 1               \\
       PDF                       & \checkmark &  4.8--6.6        & 4.5--7.8      & 3.2--7.1     & 4.6--9.5    \\
       $\mu_f$ and $\mu_r$       & \checkmark &  12.9--25.8      & 0.1--0.2      & 12.7--36.5   & 0.1--0.4     \\
       Pileup                    & \checkmark &  3.5--5.0        & 1.5--2.6      & 1.8--6.7     & 1.8--3.6     \\
       DY+Jets correction factor & \checkmark &  4.2--11.4       & ---            & 1.5--7.8     & ---           \\
       Jet energy scale          & \checkmark &  5.4--8.2        & 1.6--4.0      & 4.9--9.1     & 3.3--4.4     \\
       Jet energy resolution     & \checkmark &  2.0--3.8        & 0.6--1.8      & 3.2--6.7     & 1.7--3.8     \\
       \PV and \PH tagging       & \checkmark &  1.5--2.5        & 0.3--1.3      & 0.2--6.3     & 0.2--8.4     \\
       \cPqt\ tagging            & \checkmark &  0.5--3.0        & 4.8--7.6      & 0.2--6.3     & 0.2--8.4     \\
       misidentification of \PV  & \checkmark &  0.6--2.3        & 0.1--0.2      & 0.3--4.9     & 0.0--5.3     \\
       misidentification of \PH  & \checkmark &  0.0--0.7        & 0.0--0.7      & 0.0--14.4    & 0.0--14.4    \\
       misidentification of \cPqt& \checkmark &  1.0--2.3        & 0.2--0.4      & 6.8          & 6.8    \\
       \cPqb\ tagging            & \checkmark &  4.1--6.2        & 1.0--7.2      & 8.3--23.6    & 1.8--10.2    \\
      \hline
    \end{tabular}}
\end{table*}

The systematic uncertainties in the SM background rates are due to the uncertainties in the CMS measurements of $\rd\sigma/\rd\HT$ for \Zjets~\cite{SMP-16-015}, $\rd\sigma/\rd m_{\ttbar}$ for \ttjets~\cite{TOP-16-008}, and $\rd\sigma/\rd\pt(\PZ)$ for diboson production~\cite{SMP-16-017}. They are estimated to be 15\% in each case. The measured integrated luminosity uncertainty of 2.5\%~\cite{CMS-PAS-LUM-17-001} affects both the signal and background rate predictions. The uncertainties associated with the measured data-to-simulation efficiency scale factors for the lepton identification and the trigger efficiencies are 3 and 1\%, respectively.

{\tolerance=1200
The effect on the signal and background acceptance uncertainties due to the renormalization and factorization scale ($\mu_f$ and $\mu_r$) uncertainties and the PDF choices in the simulations are taken into account in the statistical analysis. The influence of $\mu_f$ and $\mu_r$ scale uncertainties are estimated by varying the default scales by the following six combinations of factors, ($\mu_f$, $\mu_r$) $\times$ (1/2, 1/2), (1/2, 1), (1, 1/2), (2, 2), (2, 1), and (1, 2). The maximum and minimum of the six variations are computed for each bin of the \ST distribution, producing an uncertainty ``envelope''. The uncertainties due to the PDF choices in the simulations are estimated using the {PDF4LHC} procedure~\cite{Butterworth:2015oua,Dulat:2015mca,Harland-Lang:2014zoa,Ball:2014uwa}, where the root-mean-square of 100 pseudo-experiments provided by the PDF sets represents the uncertainty envelope. The background and signal event counts are then varied relative to their nominal values up and down by a factor of two times the uncertainty envelopes. The impacts of these variations on the background and signal shape are also taken into account. The effect of the $\mu_f$ and $\mu_r$ scale uncertainties on the \TTbar and \BBbar signal yield is $<1$\%. However, this has the largest effect, amounting to as much as 36\% on the background yield. The effect due to PDF choices amounts to a 3.2--9.5\% change in the signal and background yields. The effect of the uncertainty in the pileup determination is estimated by varying the nominal {\Pp}{\Pp} inelastic cross section by 4.6\%~\cite{LHCHXSecWG}, which has an impact of 1.5--3.6\% on the signal yields. Differences between simulation and data in the jet multiplicity distributions in DY+jets background events, derived in the CR0b region as shown in Table~\ref{tab:nak4w}, are taken as an estimate of the associated systematic uncertainty, which ranges from 4.0--11.5\%.}

Several uncertainties are associated with the measurement of jet-related quantities. The jet energy scale and resolution uncertainties are about 1\%~\cite{Khachatryan:2016kdb,CMS-DP-2016-020}. The AK8 pruned jet mass scale and resolution uncertainties are evaluated to be 2.3 and 18\%~\cite{JME-16-003}, respectively. The effect of these uncertainties on the \TTbar and \BBbar signal yields is 1.5--4.4\% and 1.0--3.8\%, respectively. These uncertainties, in addition to the uncertainties in the \nsubqq (8\%) and \nsubqqq (11\%) selections~\cite{JME-16-003}, are applied for the \PV-, \PH-, and \cPqt-tagged jets. The systematic uncertainties due to the jet shower profile differences between the jets in the \Wqq and \Hbb processes are estimated from the difference observed between results obtained with the \PYTHIA~8 and \HERWIG{++} generators and are applied to the \PV- and \PH-tagged jets. The overall effect of \PV, \PH, and \cPqt\ tagging uncertainties on \TTbar and \BBbar signal yields is 0.2--8.4\%. The uncertainties in the misidentification rates of boosted jets are 5, 14, and 7\% for the \PW-, \PH-, and \cPqt-tagged jets, respectively. They are used to derive the uncertainties in the estimates of the numbers of mistagged jets in the signal and background simulated events, which result in uncertainties in the \BBbar signal yields of up to 14\%. The uncertainties in the \cPqb\ tagging efficiency scale factors are propagated to the final result, with the uncertainties in the \cPqb- and \cPqc-flavored quark jets treated as fully correlated. These uncertainties are in the range 2--5\% for \cPqb-flavored jets, a factor of two larger for \cPqc-flavored jets, and $\approx$10\% for light-flavored jets. The uncertainties due to heavy- and light-flavored jets are considered uncorrelated. Table~\ref{tab:sys} summarizes the systematic uncertainties in the background and signal yields in the \TTbar and \BBbar searches. The ranges correspond to the impact on event yields due to systematic uncertainties that affect both the rates and shapes across all the \TTbar groups or \BBbar categories. Here the \TTbar and \BBbar signals correspond to the benchmark decay channels tZtZ and bZbZ, respectively, for a \T and \B quark mass $m_{\T}=m_{\B}=1200\GeV$.

\section{Results\label{sec:Results}}

\subsection{\T quark search}

The number of observed events for the \TTbar production search in the A, B, C, and D event groups are given for the electron and muon channels in Tables~\ref{tab:evt_final_catele} and~\ref{tab:evt_final_catmu}, respectively, along with the numbers of predicted background events. The expected numbers of signal events for \T quark masses of 800 and 1200\GeV are also shown in the same tables, for three different decay scenarios, with branching fractions $\BR(\TtotZ) = 100\%$ (tZtZ), $\BR(\TtotZ) = \BR(\TtotH) = 50\%$ (tZtH), and $\BR(\TtotZ) = \BR(\TtobW) = 50\%$ (tZbW). The predicted background and observed event yields agree within their uncertainties.

\begin{table*}[!htbp]
  \centering
    \topcaption{The number of observed events and the predicted number of SM background events in the \TTbar search using \Zee channel in the four event groups. The expected numbers of signal events for \T quark masses of 800 and 1200\GeV for three different decay scenarios with assumed branching fractions $\BR(\TtotZ) = 100\%$ (tZtZ) , $\BR(\TtotZ) = \BR(\TtotH) = 50\%$ (tZtH), and $\BR(\TtotZ) = \BR(\TtobW) = 50\%$ (tZbW) are also shown. The uncertainties in the number of expected background events include the statistical and systematic uncertainties added in quadrature.}
    \label{tab:evt_final_catele}
    \begin{tabular}{p{8mm}p{23mm} D{,}{\pm}{-1} D{,}{\pm}{-1} D{,}{\pm}{-1} D{,}{\pm}{-1}}
      \hline
      \multicolumn{2}{c}{Event group} & \multicolumn{1}{c}{A}  & \multicolumn{1}{c}{B}  & \multicolumn{1}{c}{C}  & \multicolumn{1}{c}{D} \\
      \hline
      \multicolumn{2}{c}{DY+jets}            & 54.9  , 5.2  &  9.0 , 1.9  & 17.0  , 2.4  &  7.2 , 1.4 \\
      \multicolumn{2}{c}{\ttjets}            &  7.9  , 1.7  &  1.7 , 0.8  &  3.2  , 1.1  &  1.8 , 0.8 \\
      \multicolumn{2}{c}{\ttZ}               &  8.2  , 0.8  &  4.9 , 0.6  &  1.3  , 0.2  &  1.3 , 0.2 \\
      \multicolumn{2}{c}{Other backgrounds}  &  3.0  , 1.7  &  0.9 , 0.7  &  0.6  , 0.4  &  0.1 , 0.1 \\
      \multicolumn{2}{c}{Total}              & 74.1  , 6.2  & 16.5 , 2.5  & 22.2  , 2.9  & 10.4 , 1.8 \\
      \multicolumn{2}{c}{Data}               & \multicolumn{1}{c}{\text 84} & \multicolumn{1}{c}{\text 15}  & \multicolumn{1}{c}{\text 25}  & \multicolumn{1}{c}{\text 11} \\
      tZtZ,&$m_{\T}$=800\GeV                  & 54.9  , 2.2  & 43.6 , 2.0  & 9.6   , 0.9  & 9.6  , 0.9 \\
      tZtH,&$m_{\T}$=800\GeV                  & 24.8  , 1.0  & 26.7 , 0.8  & 4.2   , 0.3  & 6.5  , 0.4 \\
      tZbW,&$m_{\T}$=800\GeV                  & 24.5  , 1.0  & 17.9 , 0.6  & 5.4   , 0.3  & 5.2  , 0.3 \\
      tZtZ,&$m_{\T}$=1200\GeV                 & 3.6   , 0.1  & 3.3  , 0.1  & 0.9   , 0.1  & 0.8  , 0.1 \\
      tZtH,&$m_{\T}$=1200\GeV                 & 1.6   , 0.1  & 1.8  , 0.1  & 0.4   , 0.1  & 0.6  , 0.1 \\
      tZbW,&$m_{\T}$=1200\GeV                 & 1.6   , 0.1  & 1.3  , 0.1  & 0.5   , 0.1  & 0.4  , 0.1 \\
      \hline
    \end{tabular}

\end{table*}

\begin{table*}[!htbp]
  \centering
    \topcaption{The number of observed events and the predicted number of SM background events in the \TTbar search using $\Zmumu$ channel in the four event groups. The expected numbers of signal events for \T quark masses of 800 and 1200\GeV for three different decay scenarios with assumed branching fractions $\BR(\TtotZ) = 100\%$ (tZtZ) , $\BR(\TtotZ) = \BR(\TtotH) = 50\%$ (tZtH), and $\BR(\TtotZ) = \BR(\TtobW) = 50\%$ (tZbW) are also shown. The uncertainties in the number of expected background events include the statistical and systematic uncertainties added in quadrature.}
    \label{tab:evt_final_catmu}
    \begin{tabular}{p{8mm}p{23mm} D{,}{\pm}{-1} D{,}{\pm}{-1} D{,}{\pm}{-1} D{,}{\pm}{-1}}
      \hline
      \multicolumn{2}{c}{Event group}&  \multicolumn{1}{c}{A}  & \multicolumn{1}{c}{B}  & \multicolumn{1}{c}{C}  & \multicolumn{1}{c}{D} \\
      \hline
      \multicolumn{2}{c}{DY+jets}           & 102.5 , 10.2  & 15.8 , 3.1  & 36.8 , 4.4  & 10.2 , 2.1 \\
      \multicolumn{2}{c}{\ttjets}           &  18.4 ,  3.4  & 6.8  , 1.7  &  5.7 , 1.5  &  6.3 , 1.7 \\
      \multicolumn{2}{c}{\ttZ}              &  12.5 ,  1.2  & 7.7  , 1.0  &  2.0 , 0.3  &  2.3 , 0.3 \\
      \multicolumn{2}{c}{Other backgrounds} &   4.2 ,  1.3  & 0.9  , 0.4  &  0.5 , 0.3  &  0.3 , 0.1 \\
      \multicolumn{2}{c}{Total}             & 137.6 , 11.6  & 31.2 , 4.5  & 45.0 , 5.0  & 19.1 , 3.2 \\
      \multicolumn{2}{c}{Data}              &\multicolumn{1}{c}{\text 126} &\multicolumn{1}{c}{\text 36} &\multicolumn{1}{c}{\text 45} &\multicolumn{1}{c}{\text 22}       \\
      tZtZ,&$m_{\T}$=800\GeV            &  72.8 ,  2.5  & 65.4 , 2.4  & 10.9 , 1.0  & 11.9 , 1.0 \\
      tZtH,&$m_{\T}$=800\GeV            &  33.0 ,  0.8  & 40.0 , 0.9  &  5.5 , 0.3  &  8.4 , 0.4 \\
      tZbW,&$m_{\T}$=800\GeV            &  34.9 ,  0.9  & 26.2 , 0.8  &  7.0 , 0.4  &  7.0 , 0.4 \\
      tZtZ,&$m_{\T}$=1200\GeV           &   4.4 ,  0.1  &  3.7 , 0.1  & 1.2  , 0.1  &  1.0 , 0.1 \\
      tZtH,&$m_{\T}$=1200\GeV           &   2.0 ,  0.1  &  2.2 , 0.1  & 0.6  , 0.1  &  0.8 , 0.1 \\
      tZbW,&$m_{\T}$=1200\GeV           &   1.9 ,  0.1  &  1.4 , 0.1  & 0.7  , 0.1  &  0.5 , 0.1 \\
      \hline
    \end{tabular}

\end{table*}

\begin{figure*}[!hbtp]
  \centering
    \includegraphics[width=0.48\textwidth]{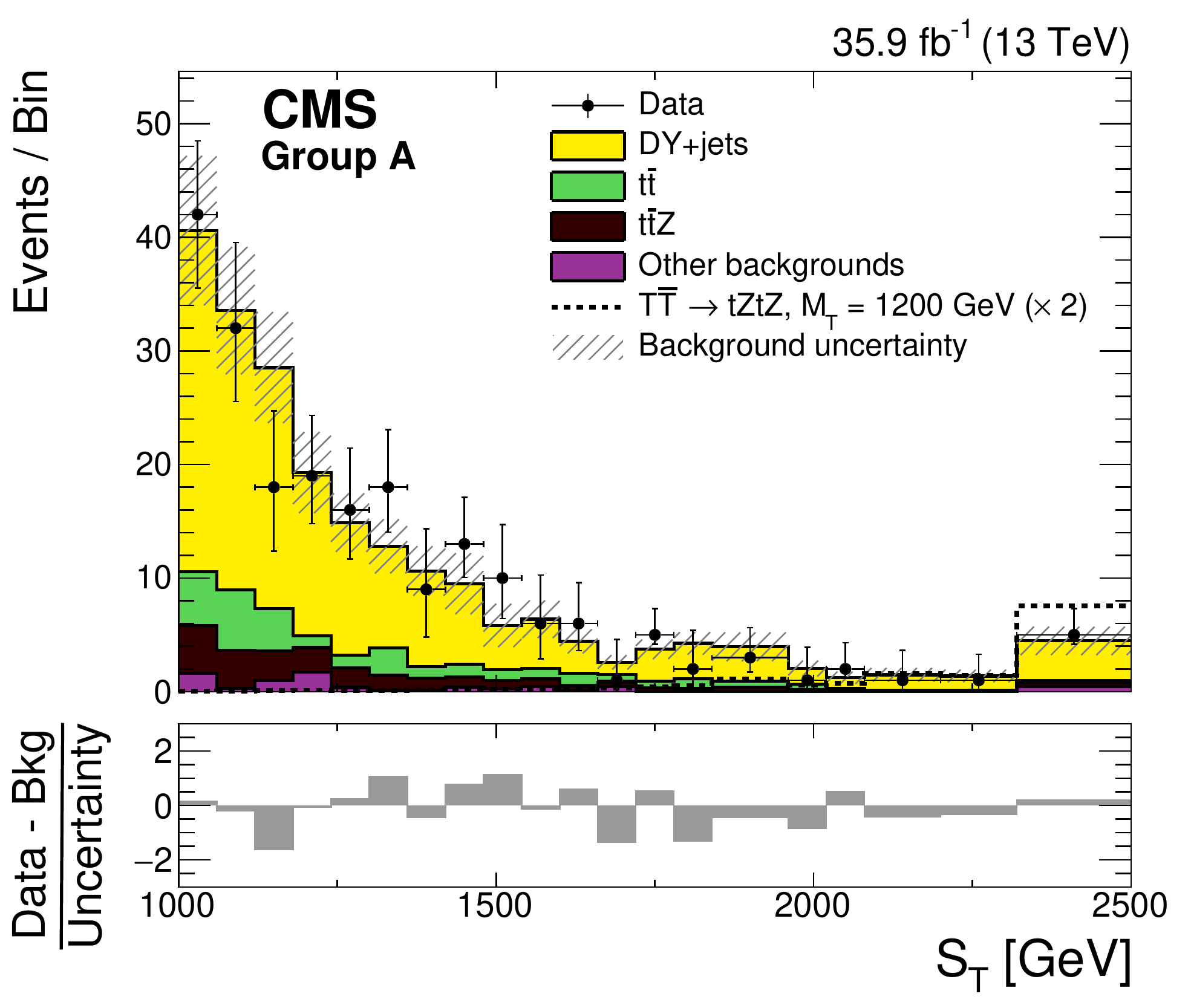}
    \includegraphics[width=0.48\textwidth]{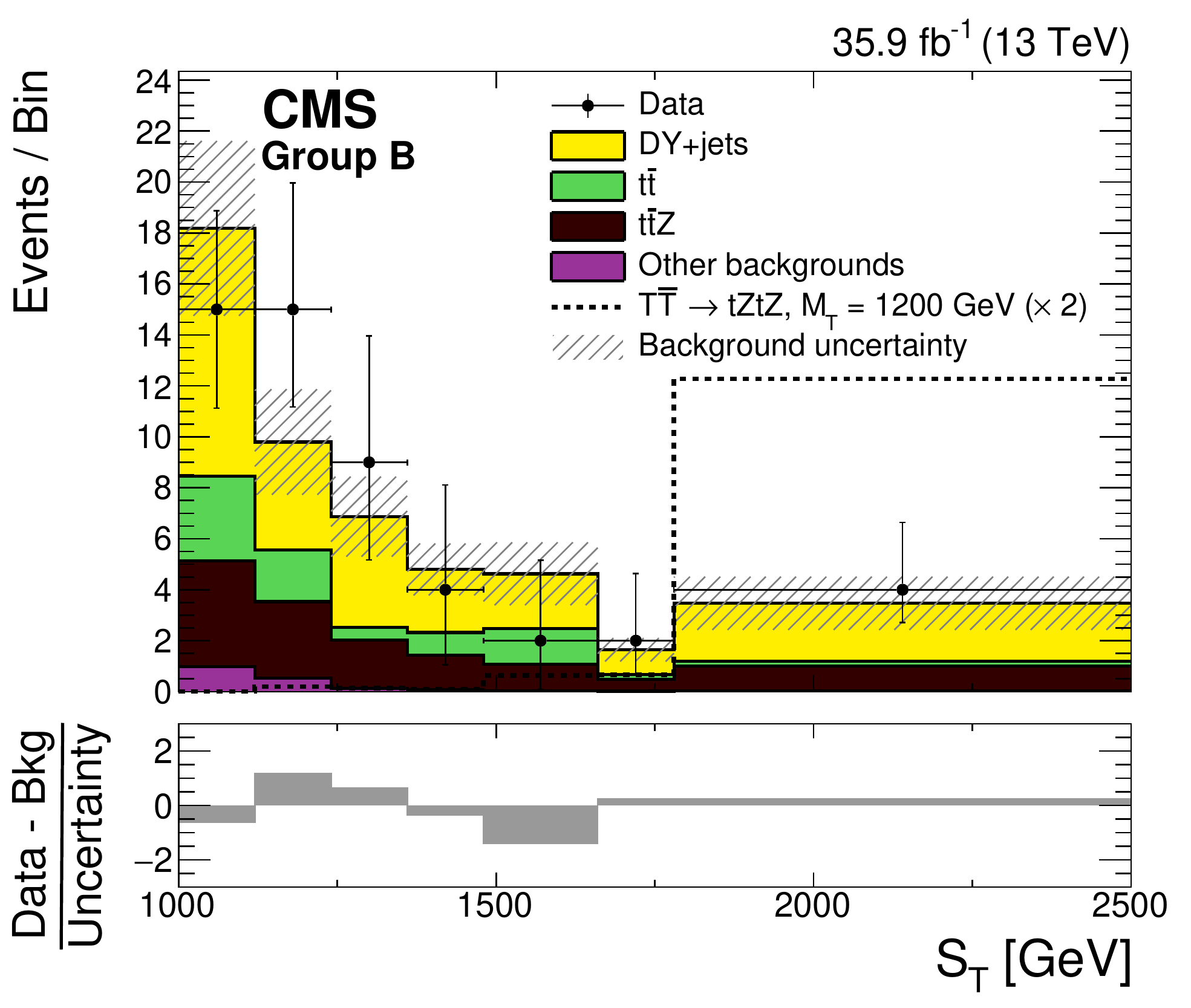}\\
    \includegraphics[width=0.48\textwidth]{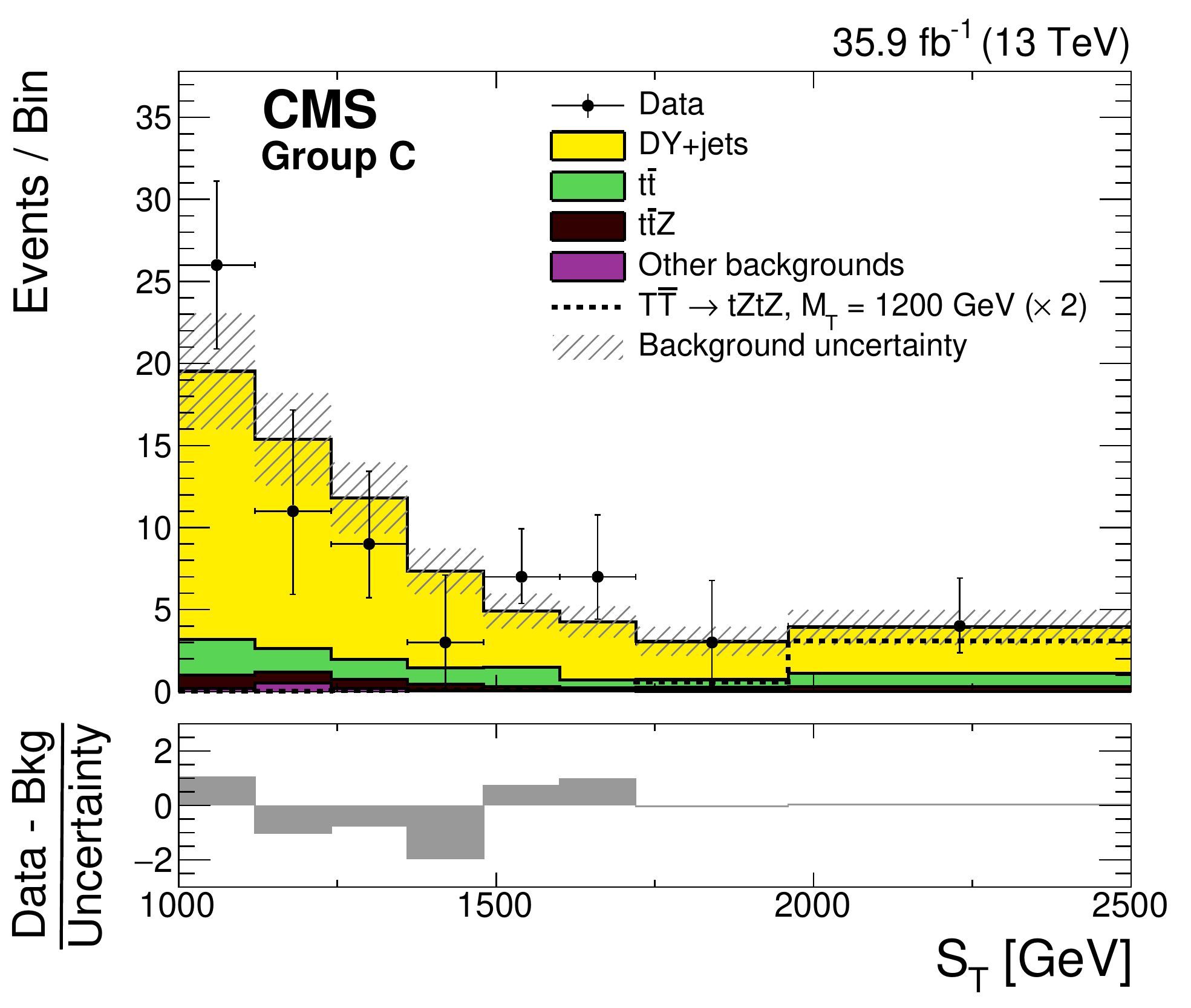}
    \includegraphics[width=0.48\textwidth]{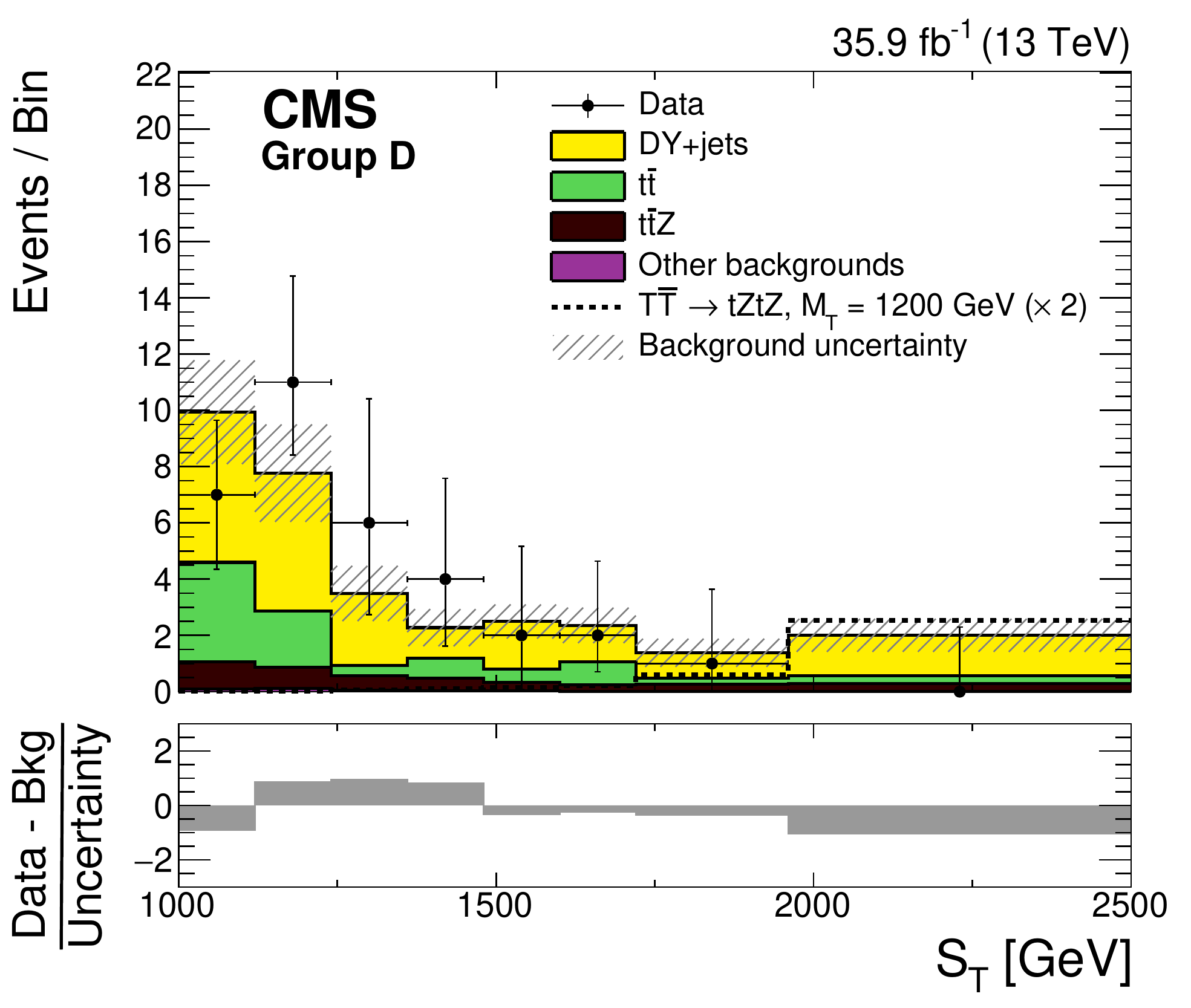}
    \caption{The \ST distributions for groups A, B, C, D (left to right, upper to lower) from data (points with vertical and horizontal bars), the expected SM backgrounds (shaded histograms), and the expected signal, scaled up by a factor 2, for \TTtotZtZ with $m_{\T}=1200\GeV$ (dotted lines). The vertical bars on the points show the central 68\% \CL intervals for Poisson-distributed data. The horizontal bars give the bin widths. The hatched bands represent the statistical and systematic uncertainties in the total background contribution added in quadrature. The lower plots give the difference between the data and the total expected background, divided by the total background uncertainty.}
    \label{fig:st_TT}

\end{figure*}

To determine the upper limits on the \TTbar cross section, the electron and muon channels are combined, and a simultaneous binned maximum-likelihood fit is performed on the \ST distributions in data for the four event groups. The measured \ST distributions in data are shown in Fig.~\ref{fig:st_TT} for each of the event groups, along with the predicted background distributions and the expected signal distributions for \TTtotZtZ with $m_{\T}=1200\GeV$. The impact of the statistical uncertainty in the simulated samples is reduced by rebinning each \ST distribution to ensure that the statistical uncertainty associated with the expected background is less than 20\% in each bin. There is no indication of a signal in the \ST distribution of any of the event groups.

{\tolerance=800
The upper limits at 95\% \CL on the \TTbar cross section are computed using a Bayesian likelihood-based technique~\cite{PDG2018} with the \textsc{Theta} framework~\cite{theta}. All the systematic uncertainties due to normalization variations described in the previous section enter the likelihood as nuisance parameters with log-normal prior distributions, whereas the uncertainties from the shape variations are assigned Gaussian-distributed priors. For the signal cross section parameter, we use a uniform prior distribution. The likelihood is marginalized with respect to the nuisance parameters, and the limits are extracted from a simultaneous maximum-likelihood fit of the \ST distributions in all four groups shown in Fig.~\ref{fig:st_TT}.
\par}

\begin{figure*}[hbtp]
  \centering
    \includegraphics[width=0.48\textwidth]{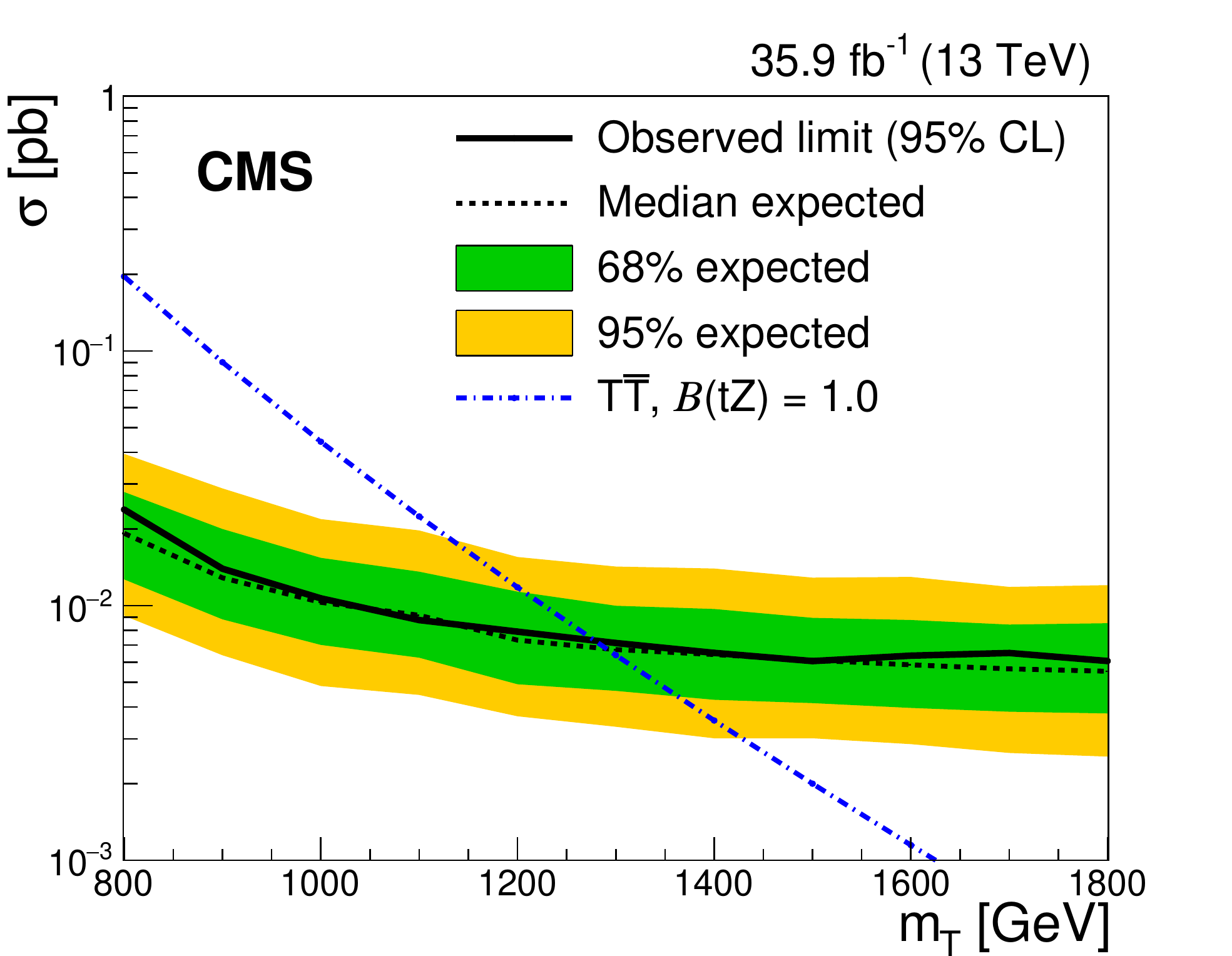}
    \includegraphics[width=0.48\textwidth]{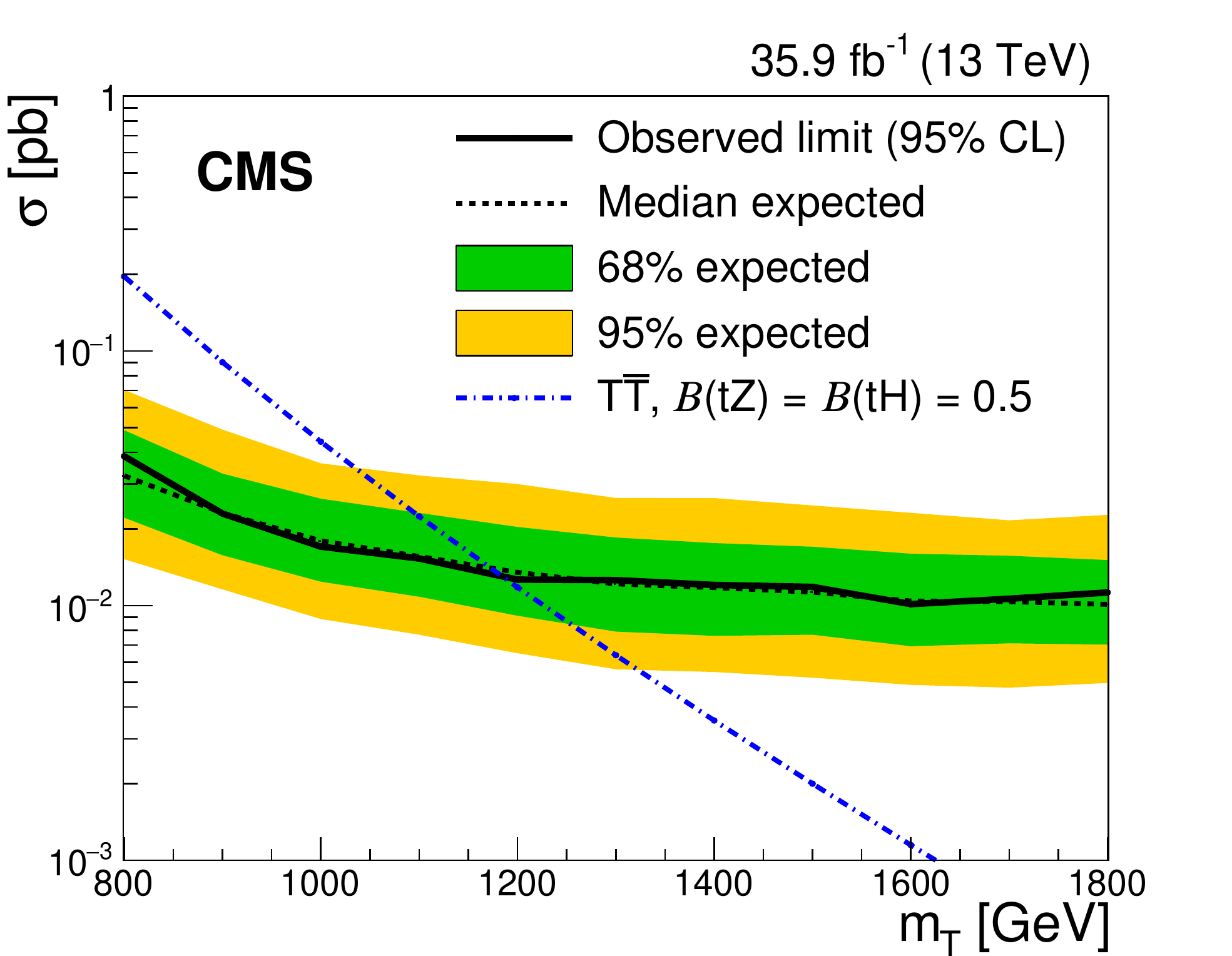}
    \includegraphics[width=0.48\textwidth]{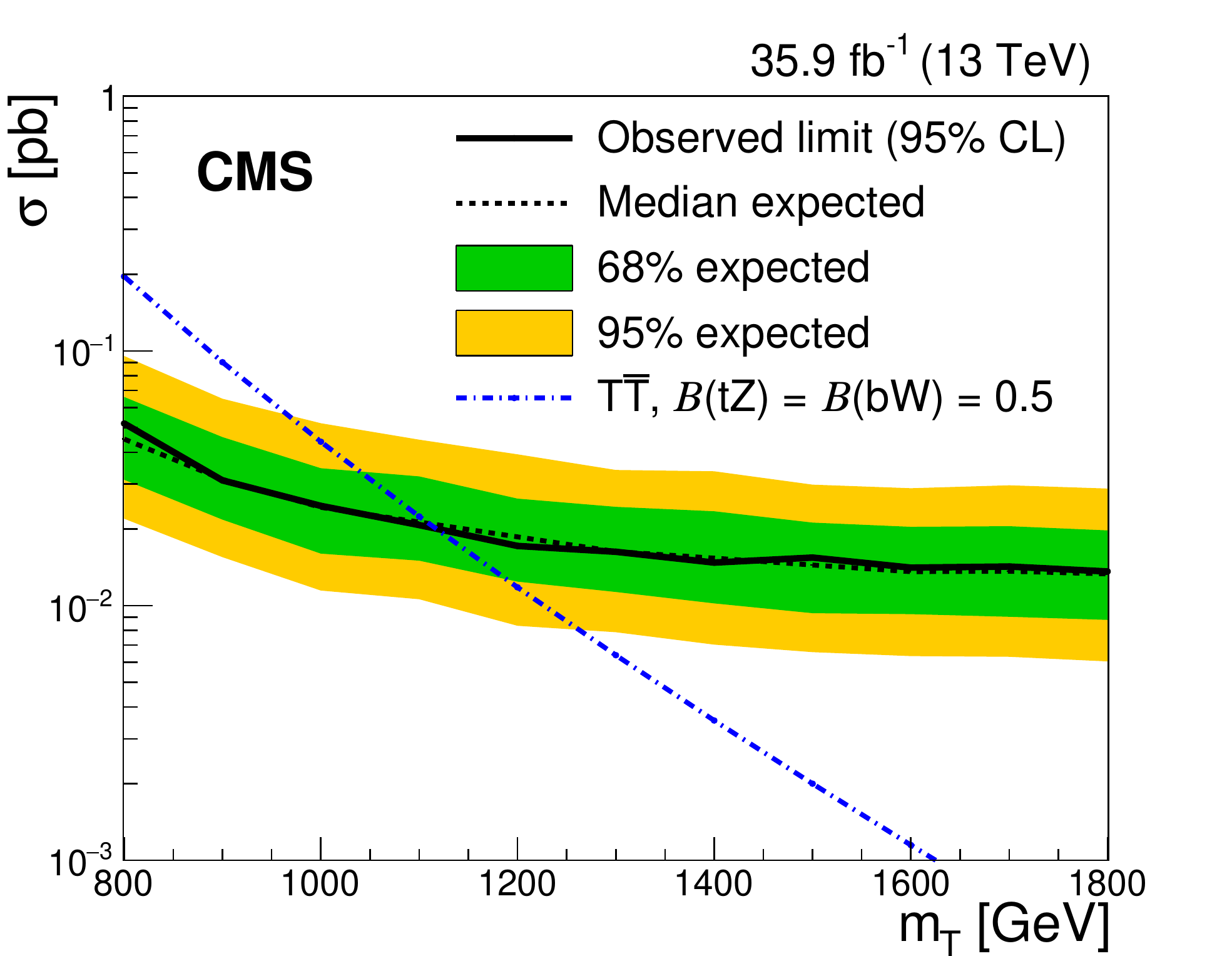}
    \caption{The observed (solid line) and expected (dashed line) 95\% \CL upper limits on the \TTbar cross section as a function of the \T quark mass assuming (upper left) $\BR(\TtotZ) = 100\%$, (upper right) $\BR(\TtotZ) = \BR(\TtotH)  = 50\%$, and (lower) $\BR(\TtotZ) = \BR(\TtobW) = 50\%$. The dotted-dashed curve displays the theoretical \TTbar production cross section. The inner and outer bands show the one and two standard deviation uncertainties in the expected limits, respectively.}
   \label{fig:exp-tZ-tH-bW}

\end{figure*}

The upper limits on the \TTbar cross section are computed for different \T quark mass values and for the three branching fraction scenarios listed above. The upper limits at 95\% \CL on the \TTbar cross section are shown as a function of the \T quark mass by the solid line in Fig.~\ref{fig:exp-tZ-tH-bW}. The median expected upper limit is given by the dotted line, while the inner and outer bands correspond to one and two standard deviation uncertainties, respectively, in the expected limit. The dotted-dashed curve displays the predicted theoretical signal cross section~\cite{Czakon:2013goa}. Comparing the observed cross section limits to the theoretical signal cross section, we exclude \T quarks with masses less than 1280, 1185, and 1120\GeV, respectively, for the three branching ratio hypotheses listed above. The expected upper limits are 1290, 1175, and 1115\GeV for the respective scenarios.

\begin{figure*}[hbtp]
  \centering
    \includegraphics[width=0.48\textwidth]{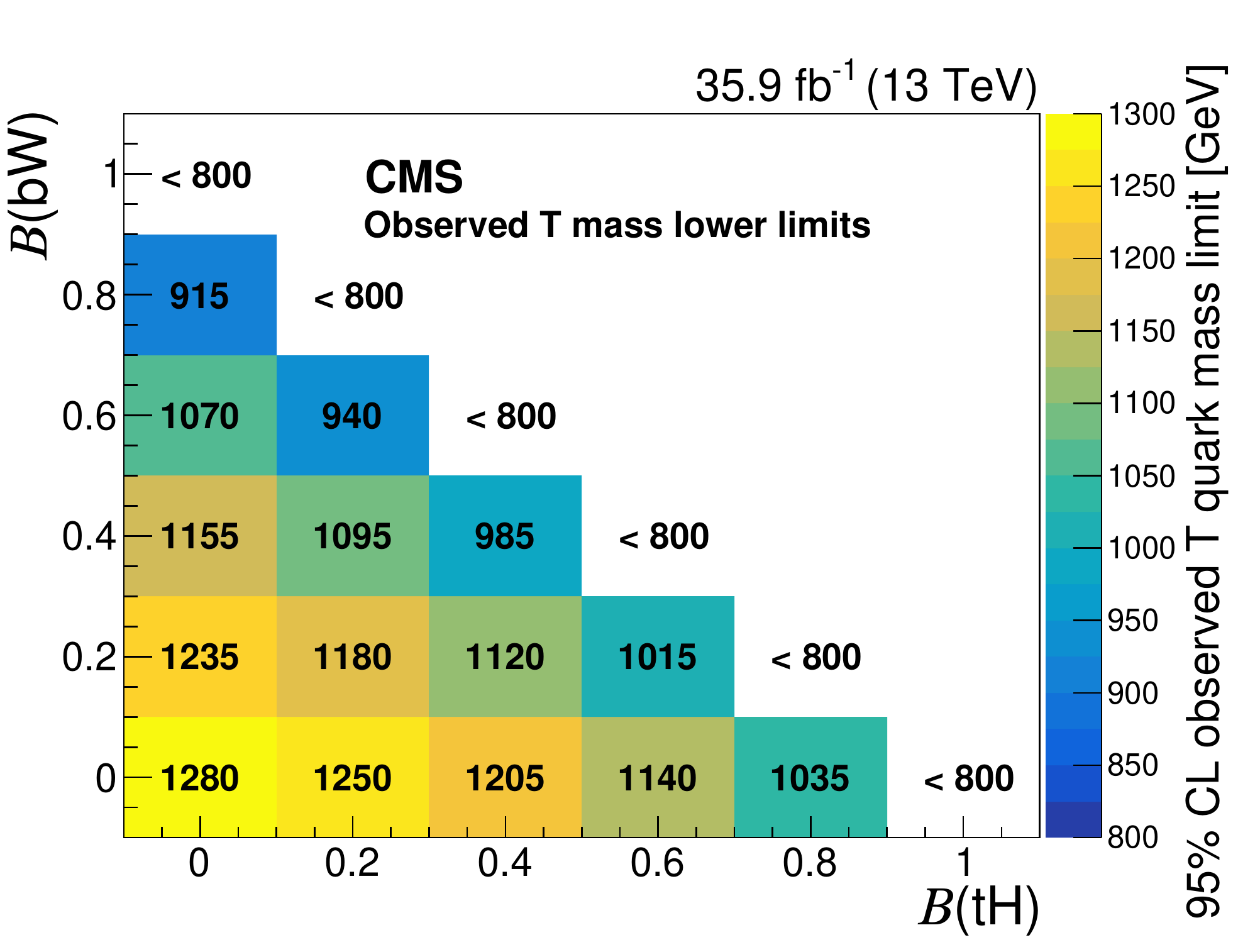}
    \includegraphics[width=0.48\textwidth]{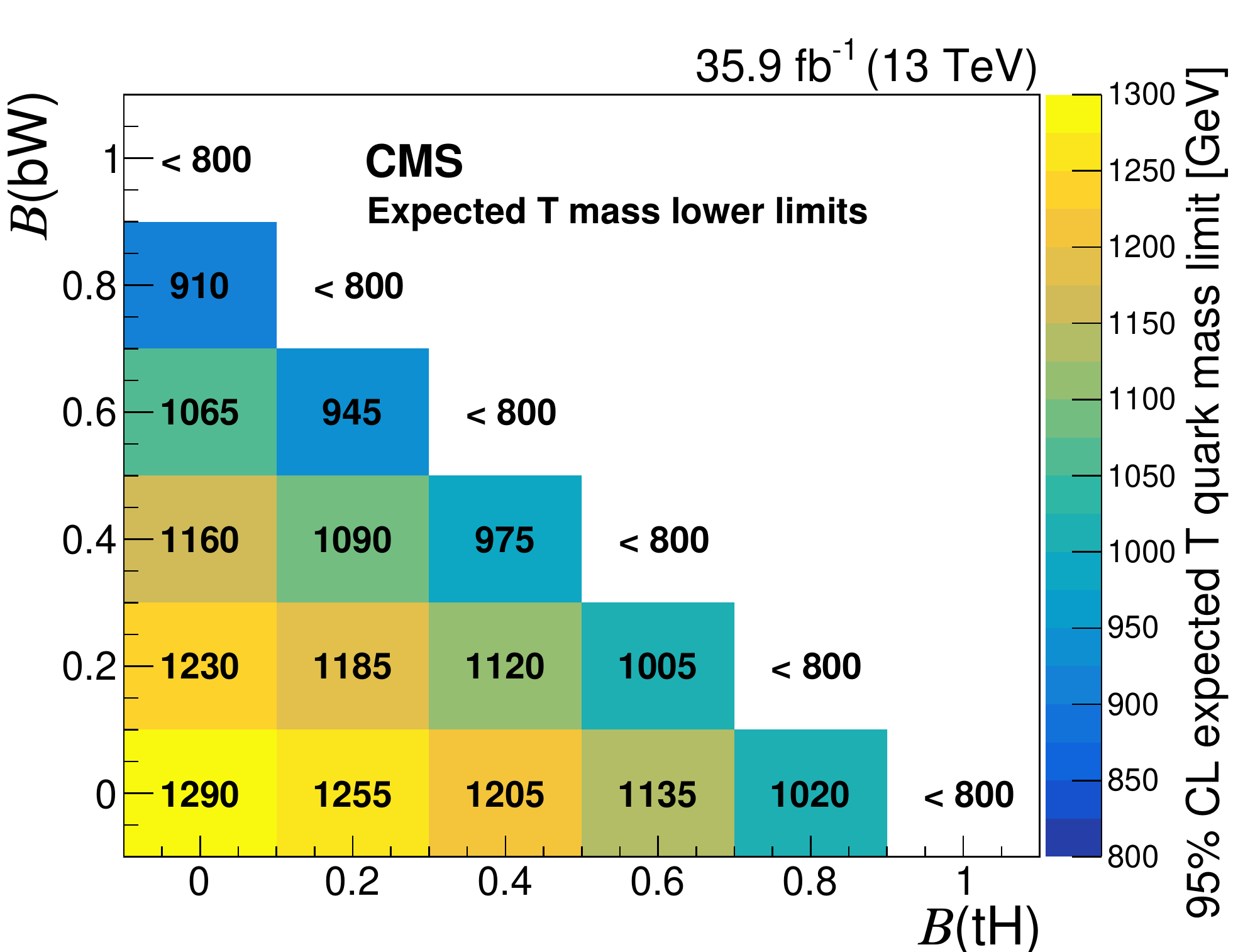} \\
    \includegraphics[width=0.48\textwidth]{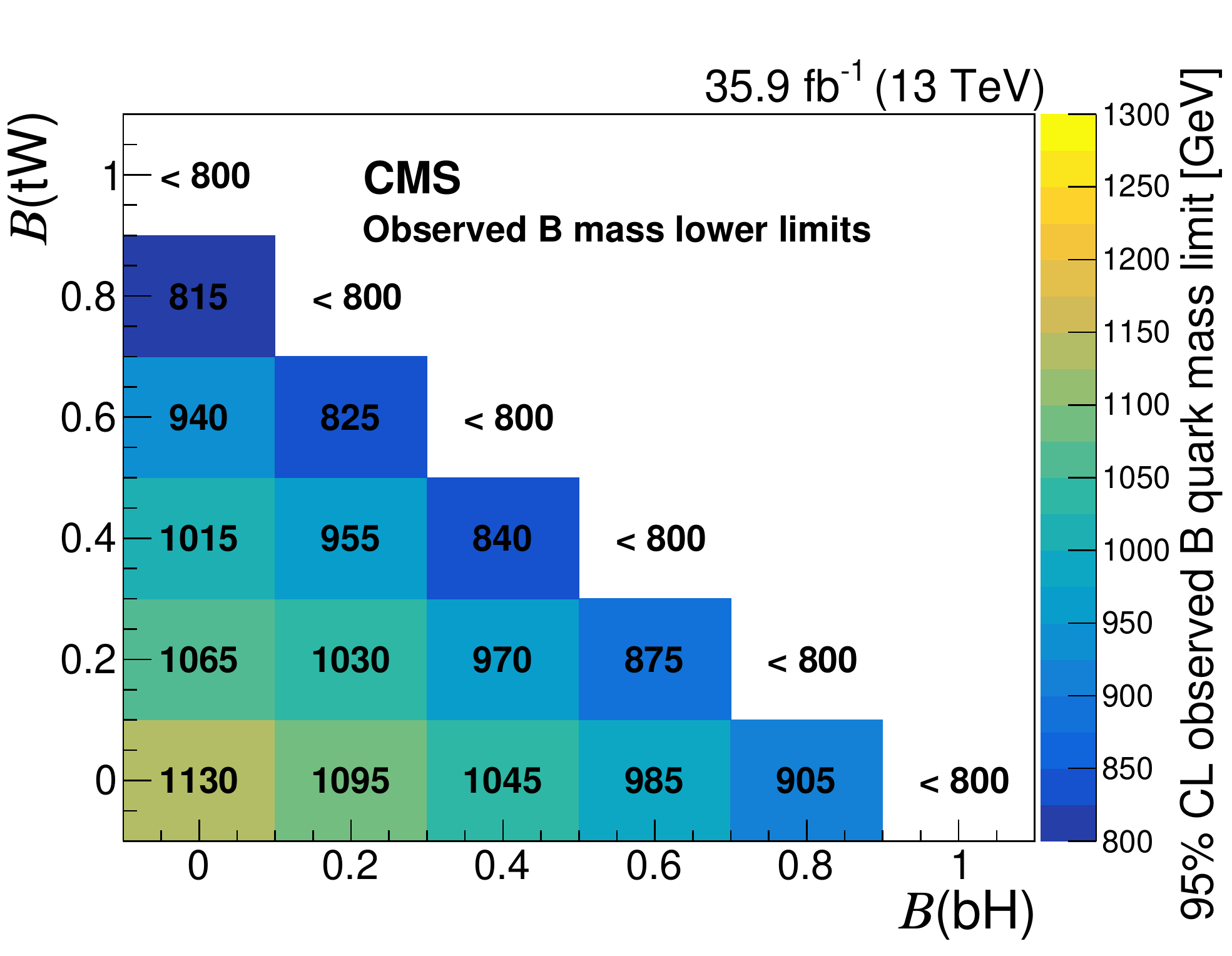}
    \includegraphics[width=0.48\textwidth]{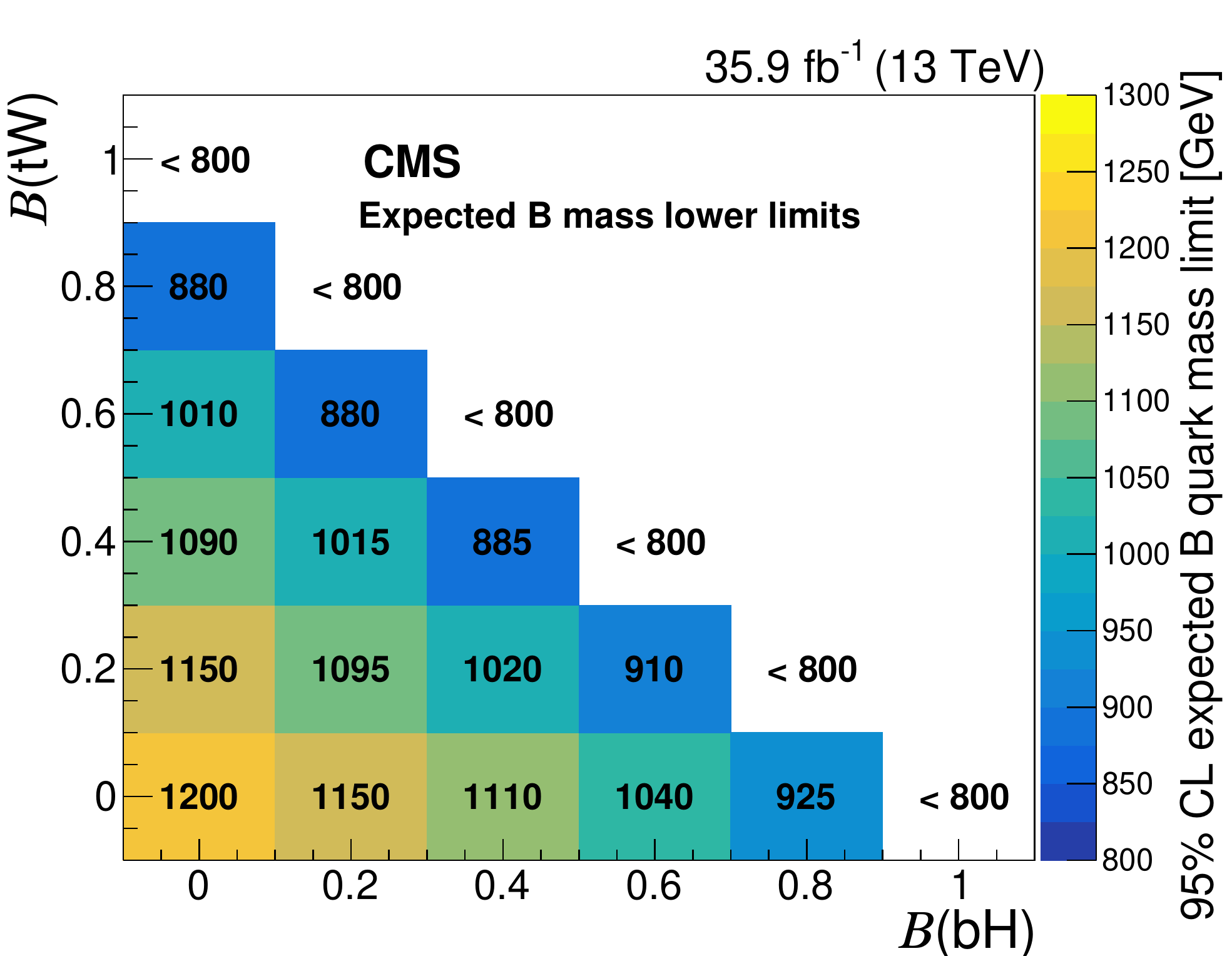}
    \caption{The observed (left) and expected (right) 95\% \CL lower limits on the mass of the \T (upper) and \B (lower) quark, in \GeV, for various branching fraction scenarios, assuming $\BR(\TtotZ) + \BR(\TtotH) + \BR(\TtobW) = 1$ and $\BR(\BtobZ) + \BR(\BtobH) + \BR(\BtotW) = 1$, respectively.}
    \label{fig:limit-All}

\end{figure*}

Figure~\ref{fig:limit-All} (upper) displays the observed (left) and expected (right) 95\% \CL lower limits on the \T quark mass as a function of the relevant branching fractions, assuming $\BR(\TtotZ) + \BR(\TtotH) + \BR(\TtobW) = 1.0$. For a \T quark decaying exclusively via \TtotZ, the lower mass limit is 1280\GeV.

\subsection{\B quark search}

\begin{figure*}[!htbp]
\centering
  \includegraphics[width=0.48\textwidth]{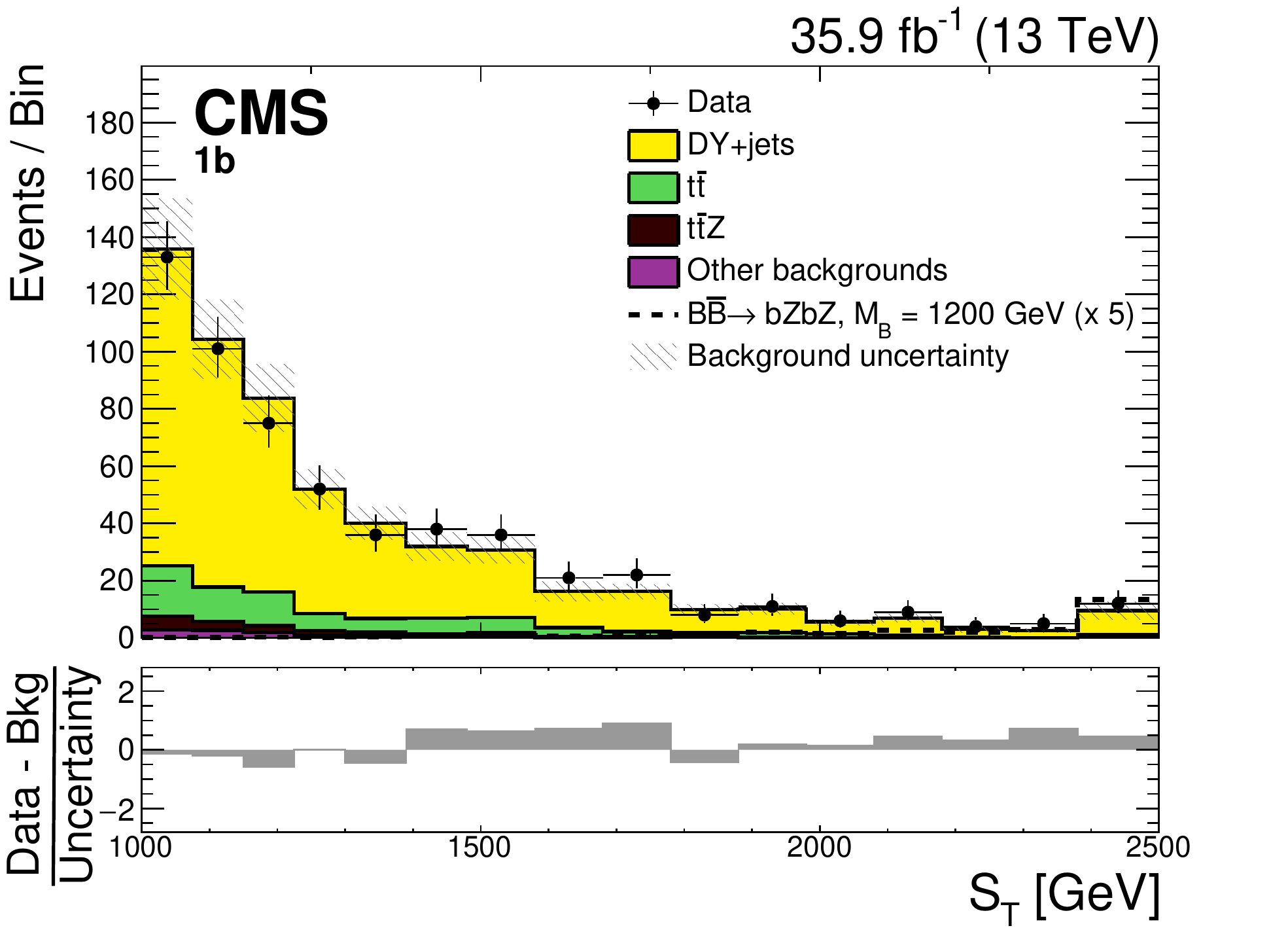}
  \includegraphics[width=0.48\textwidth]{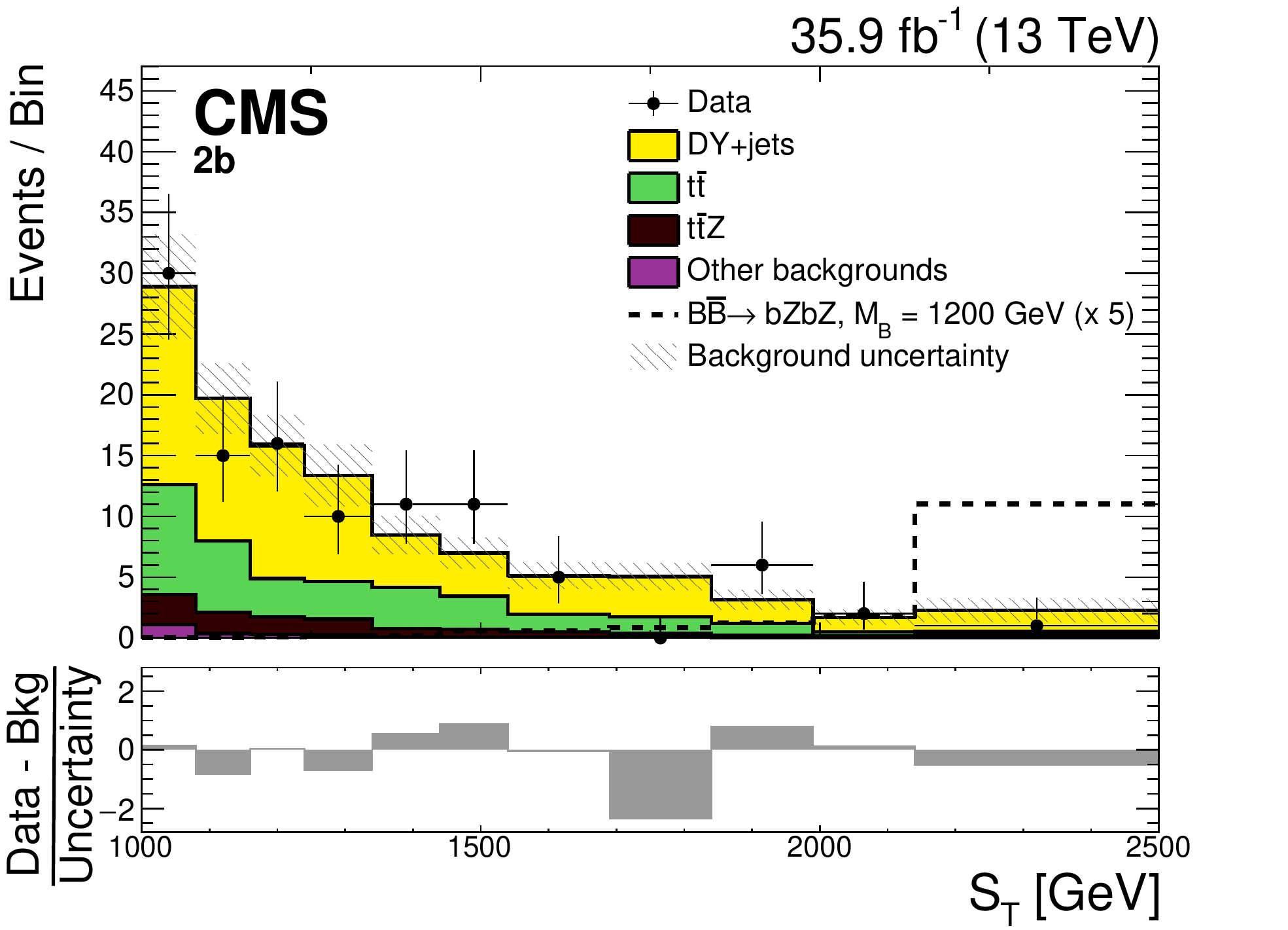}\\
  \includegraphics[width=0.48\textwidth]{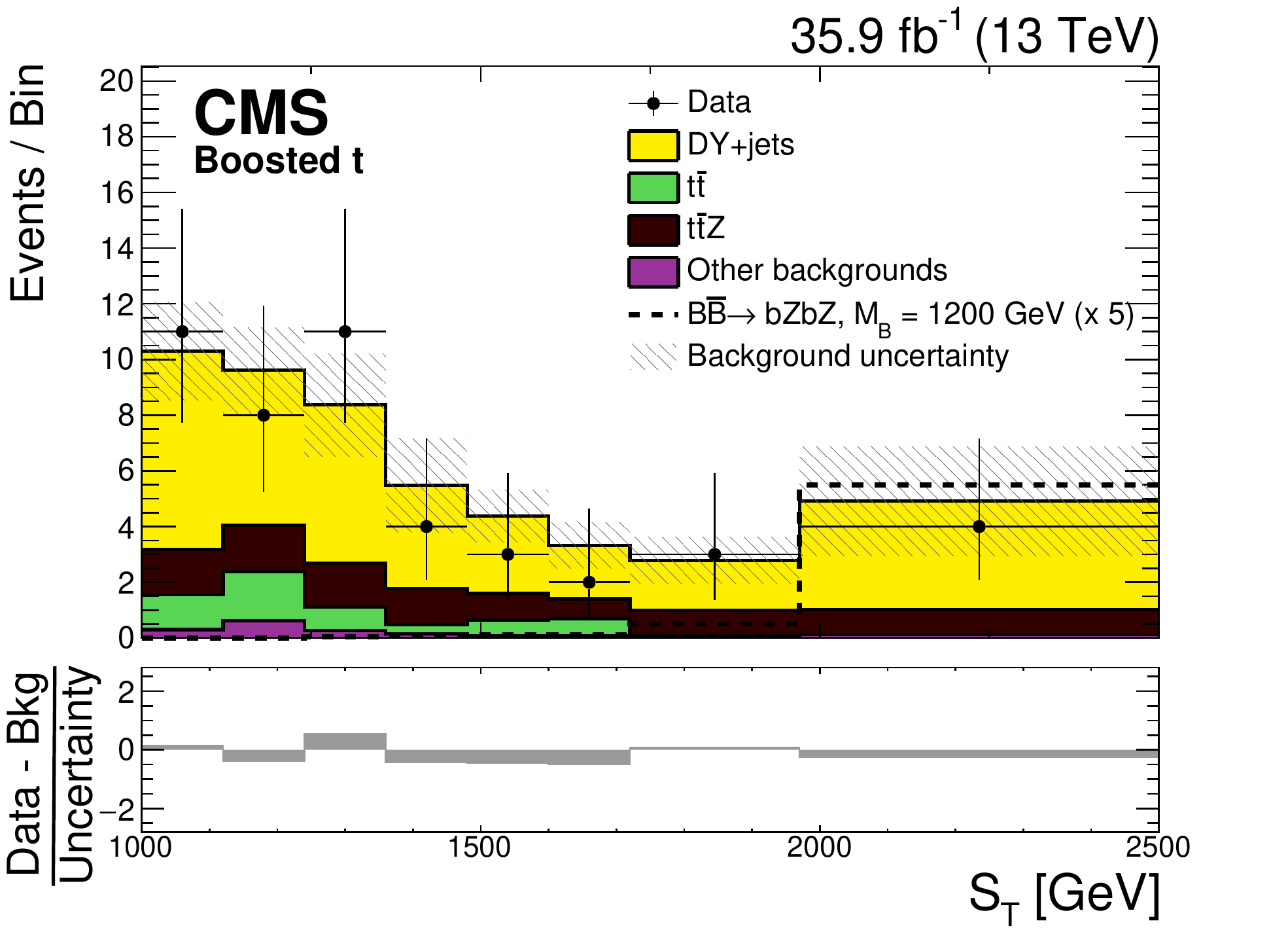}
  \includegraphics[width=0.48\textwidth]{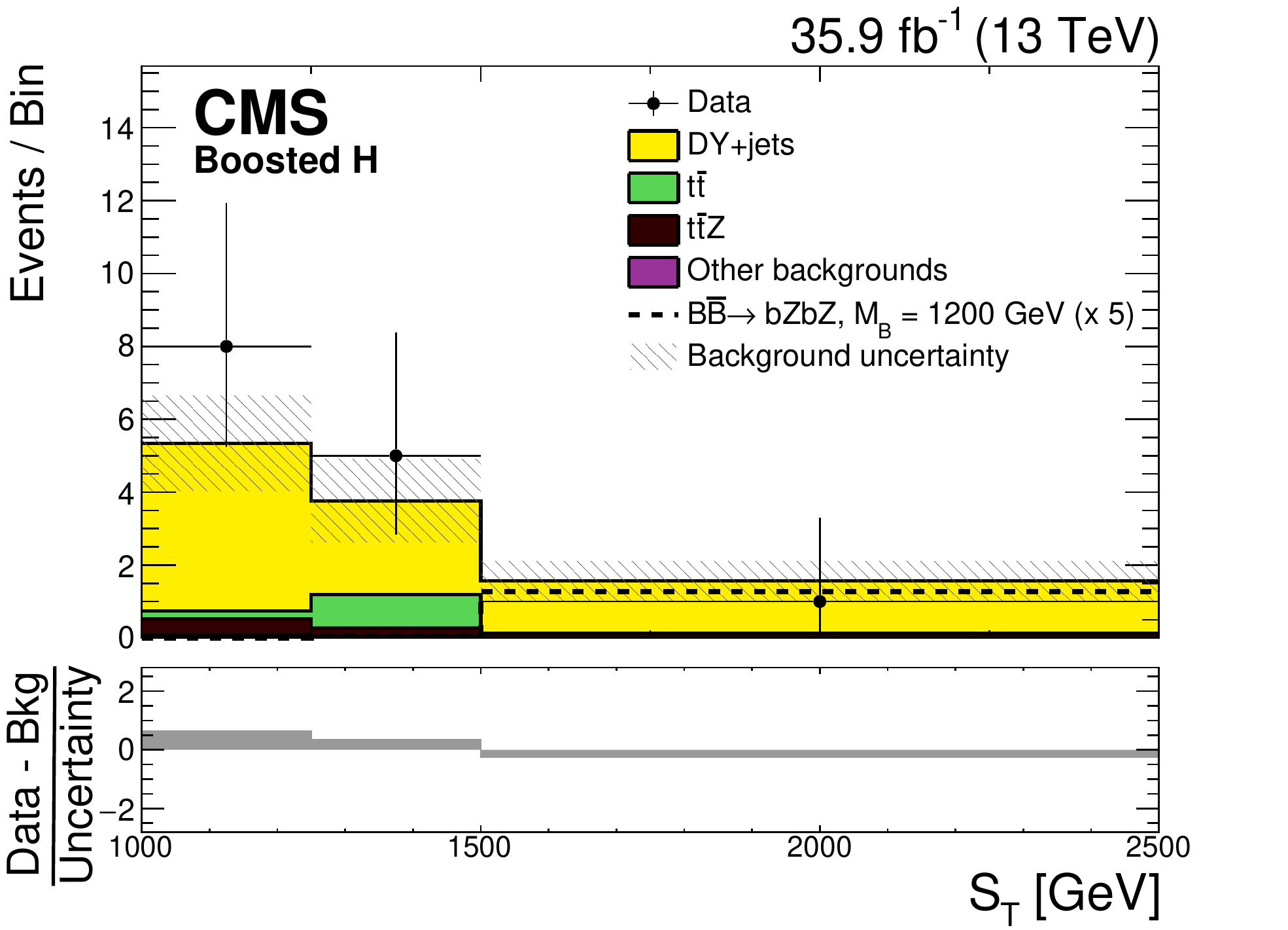} \\
  \includegraphics[width=0.48\textwidth]{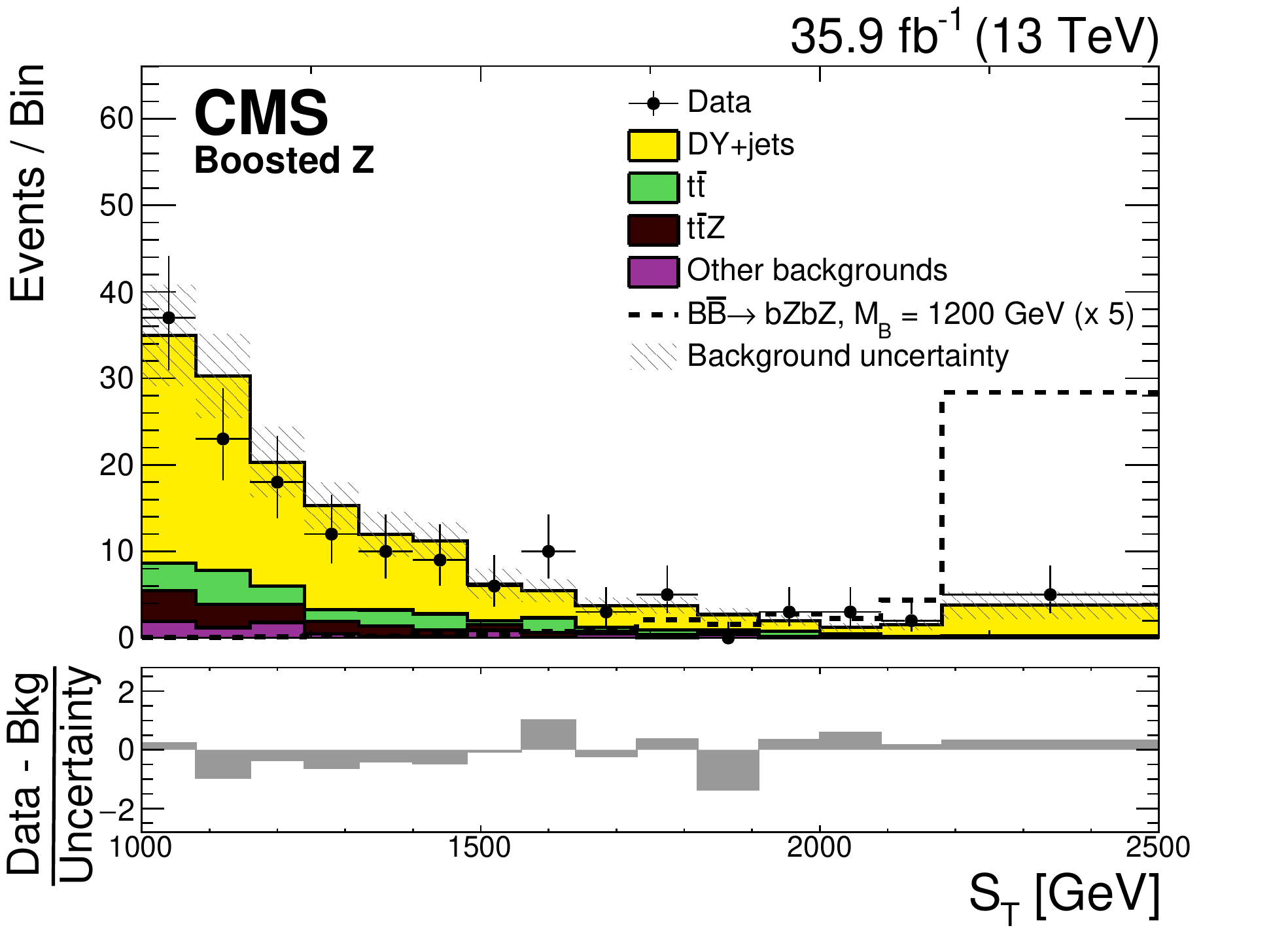} \\
      \caption{The \ST distributions for the 1\cPqb, 2\cPqb, boosted \cPqt, boosted \PH and boosted \PZ (left to right, upper to lower) event categories for the data (points with vertical and horizontal bars), and the expected background (shaded histograms). The vertical bars give the statistical uncertainty in the data, and the horizontal bars show the bin widths. The expected signal for \BBtobZbZ with $m_{\B} = 1200\GeV$ multiplied by a factor of 5 is shown by the dashed line. The statistical and systematic uncertainties in the SM background prediction, added in quadrature, are represented by the hatched bands. The lower panel in each plot show the difference between the data and the expected background, divided by the total uncertainty.}
  \label{fig:BpBpTobZbZ}

\end{figure*}

The numbers of observed and predicted background events in the five event categories for the \BBbar search using $\Zee$ and $\Zmumu$ are given in Tables~\ref{tab:evt_BB_ele} and~\ref{tab:evt_BB_muon}, respectively. The expected number of signal events in each category is also shown for \B masses of 800 and 1200\GeV. The branching fraction hypotheses assumed for the three decay channels are $\BR(\BtobZ) = 100\%$ (bZbZ), $\BR(\BtobZ) = \BR(\BtobH) = 50\%$ (bZbH), and $\BR(\BtobZ) = \BR(\BtotW) = 50\%$ (bZtW). The numbers of observed and expected background events are consistent with each other for every event category. As with the \TTbar search, 95\% \CL upper limits on the \BBbar production cross section are determined using a simultaneous binned maximum-likelihood fit to the \ST distributions for the different event categories, shown in Fig.~\ref{fig:BpBpTobZbZ}.

\begin{table*}[!htbp]
  \centering
    \topcaption{The numbers of observed events and the predicted number of SM background events in the \BBbar search for the five event categories using \Zee channel. The expected numbers of signal events for \B masses of 800 and 1200\GeV with branching fraction hypotheses for the three decay channels, $\BR(\BtobZ) = 100\%$ (bZbZ), $\BR(\BtobZ) = \BR(\BtobH) = 50\%$ (bZbH), and $\BR(\BtobZ) = \BR(\BtotW) = 50\%$ (bZtW) are also shown. The uncertainties in the number of expected background events include the statistical and systematic uncertainties added in quadrature.}
    \label{tab:evt_BB_ele}
    \begin{tabular}{p{8mm}p{23mm} D{,}{\pm}{-1} D{,}{\pm}{-1} D{,}{\pm}{-1} D{,}{\pm}{-1} D{,}{\pm}{-1}}
      \hline
      \multicolumn{2}{c}{Event category}& \multicolumn{1}{c}{1\cPqb}        &  \multicolumn{1}{c}{2\cPqb}      & \multicolumn{1}{c}{Boosted \cPqt}       & \multicolumn{1}{c}{Boosted \PH}   & \multicolumn{1}{c}{Boosted \PV}  \\

      \hline
      \multicolumn{2}{c}{DY+jets}                         & 155.2 , 10.4  & 23.5 , 3.2  & 9.5 , 1.8  & 1.9 ,  1.0  & 37.8 ,  4.4 \\
      \multicolumn{2}{c}{\ttjets}                         & 16.7  , 3.1   & 6.9  , 2.1  & 0.5 , 0.6  & 0.3 ,  0.6  & 5.1  ,  1.8 \\
      \multicolumn{2}{c}{\ttZ}                            & 6.0   , 0.7   & 3.4  , 0.5  & 3.3 , 0.5  & 0.0 ,  0.4  & 5.2  ,  0.6 \\
      \multicolumn{2}{c}{Other backgrounds}               & 6.7   , 3.8   & 1.3  , 1.3  & 0.9 , 0.6  & 0.0 ,  0.4  & 3.6  ,  2.5 \\
      \multicolumn{2}{c}{Total}                           & 184.6 , 12.7  & 35.1 , 4.2  & 14.2 , 2.1  & 2.5 , 1.1  & 51.7 , 5.3 \\
      \multicolumn{2}{c}{Data}                            &\multicolumn{1}{c}{\text 192} &\multicolumn{1}{c}{\text 37} &\multicolumn{1}{c}{\text 19} &\multicolumn{1}{c}{\text 6} &\multicolumn{1}{c}{\text 54}              \\
      bZbZ,&$m_{\B}$=800\GeV           & 39.3  ,  1.8  &   24.6 , 1.4  & 7.3 , 0.8   & 2.1 , 0.4   & 58.2 , 2.3  \\
      bZbH,&$m_{\B}$=800\GeV           & 20.5  ,  0.7  &   18.2 , 0.6  & 4.7 , 0.3   & 4.6 , 0.3   & 23.3 , 0.7  \\
      bZtW,&$m_{\B}$=800\GeV           & 18.8  ,  0.6  &   11.5 , 0.5  & 7.1 , 0.4   & 1.0 , 0.2   & 29.9 , 0.8  \\
      bZbZ,&$m_{\B}$=1200\GeV          & 2.6   ,  0.1  &   1.3  , 0.1  & 0.6 , 0.1   & 0.2 , 0.1   & 3.9  , 0.2  \\
      bZbH,&$m_{\B}$=1200\GeV          & 1.4   ,  0.1  &   1.1  , 0.1  & 0.4 , 0.1   & 0.4 , 0.1   & 1.6  , 0.1  \\
      bZtW,&$m_{\B}$=1200\GeV          & 1.2   ,  0.1  &   0.6  , 0.1  & 0.7 , 0.1   & 0.1 , 0.1   & 1.9  , 0.1  \\
      \hline
    \end{tabular}

\end{table*}

\begin{table*}[!htbp]
  \centering
      \topcaption{The number of observed events and the predicted number of SM background events in the \BBbar search for the five event categories using \Zmumu channel. The expected numbers of signal events for \B masses of 800 and 1200\GeV with branching fraction hypotheses for the three decay channels, $\BR(\BtobZ) = 100\%$ (bZbZ), $\BR(\BtobZ) = \BR(\BtobH) = 50\%$ (bZbH), and $\BR(\BtobZ) = \BR(\BtotW) = 50\%$ (bZtW) are also shown. The uncertainties in the number of expected background events include the statistical and systematic uncertainties added in quadrature.}
    \label{tab:evt_BB_muon}
    \begin{tabular}{p{8mm}p{23mm} D{,}{\pm}{-1} D{,}{\pm}{-1} D{,}{\pm}{-1} D{,}{\pm}{-1} D{,}{\pm}{-1}}
    \hline
    \multicolumn{2}{c}{Event category}                 & \multicolumn{1}{c}{1\cPqb}       &  \multicolumn{1}{c}{2\cPqb}     & \multicolumn{1}{c}{Boosted \cPqt}      & \multicolumn{1}{c}{Boosted \PH}  & \multicolumn{1}{c}{Boosted \PV}  \\
    \hline
    \multicolumn{2}{c}{DY+jets}                        & 280.6 , 20.2 & 38.1 , 4.6 & 19.8 , 3.2  & 5.0 , 1.6  & 71.5  , 7.6  \\
    \multicolumn{2}{c}{\ttjets}                        & 45.1  , 5.6  & 20.0 , 3.4 & 3.9  , 1.3  & 0.6 , 0.8  & 10.8  , 2.9  \\
    \multicolumn{2}{c}{\ttZ}                           & 9.0   , 0.9  & 5.3  , 0.6 & 5.4  , 0.6  & 0.4 , 0.4  & 8.0   , 0.8  \\
    \multicolumn{2}{c}{Other backgrounds}              & 6.1   , 4.2  & 1.2  , 0.6 & 0.9  , 0.5  & 0.1 , 0.4  & 4.5   , 3.1  \\
    \multicolumn{2}{c}{Total}                          & 340.7 , 22.3  & 64.5 , 6.4 & 30.0 , 3.7 & 6.1 , 1.8 & 94.7 , 9.1  \\
    \multicolumn{2}{c}{Data}                           &\multicolumn{1}{c}{\text 374} &\multicolumn{1}{c}{\text 70} &\multicolumn{1}{c}{\text 27} &\multicolumn{1}{c}{\text 8} &\multicolumn{1}{c}{\text 92}\\

    bZbZ,&$m_{\B}$=800\GeV      & 56.7 , 2.1  & 38.8 ,  1.8 & 8.7 ,  0.9  & 2.3 ,  0.4  & 73.3 ,  2.6  \\
    bZbH,&$m_{\B}$=800\GeV      & 27.9 , 0.8  & 27.5 ,  0.8 & 6.8 ,  0.4  & 6.7 ,  0.4  & 30.2 ,  0.8  \\
    bZtW,&$m_{\B}$=800\GeV      & 26.3 , 0.7  & 16.2 ,  0.6 & 9.4 ,  0.5  & 1.2 ,  0.2  & 38.6 ,  0.9  \\
    bZbZ,&$m_{\B}$=1200\GeV     & 3.3  , 0.1  & 1.9  ,  0.1 & 0.7 ,  0.1  & 0.1 ,  0.1  & 4.8  ,  0.2  \\
    bZbH,&$m_{\B}$=1200\GeV     & 1.7  , 0.1  & 1.3  ,  0.1 & 0.5 ,  0.1  & 0.5 ,  0.1  & 2.0  ,  0.1  \\
    bZtW,&$m_{\B}$=1200\GeV     & 1.5  , 0.1  & 0.8  ,  0.1 & 0.8 ,  0.1  & 0.1 ,  0.1  & 2.4  ,  0.1  \\
    \hline
    \end{tabular}

\end{table*}

\begin{figure*}[hbtp]
  \centering
    \includegraphics[width=0.48\textwidth]{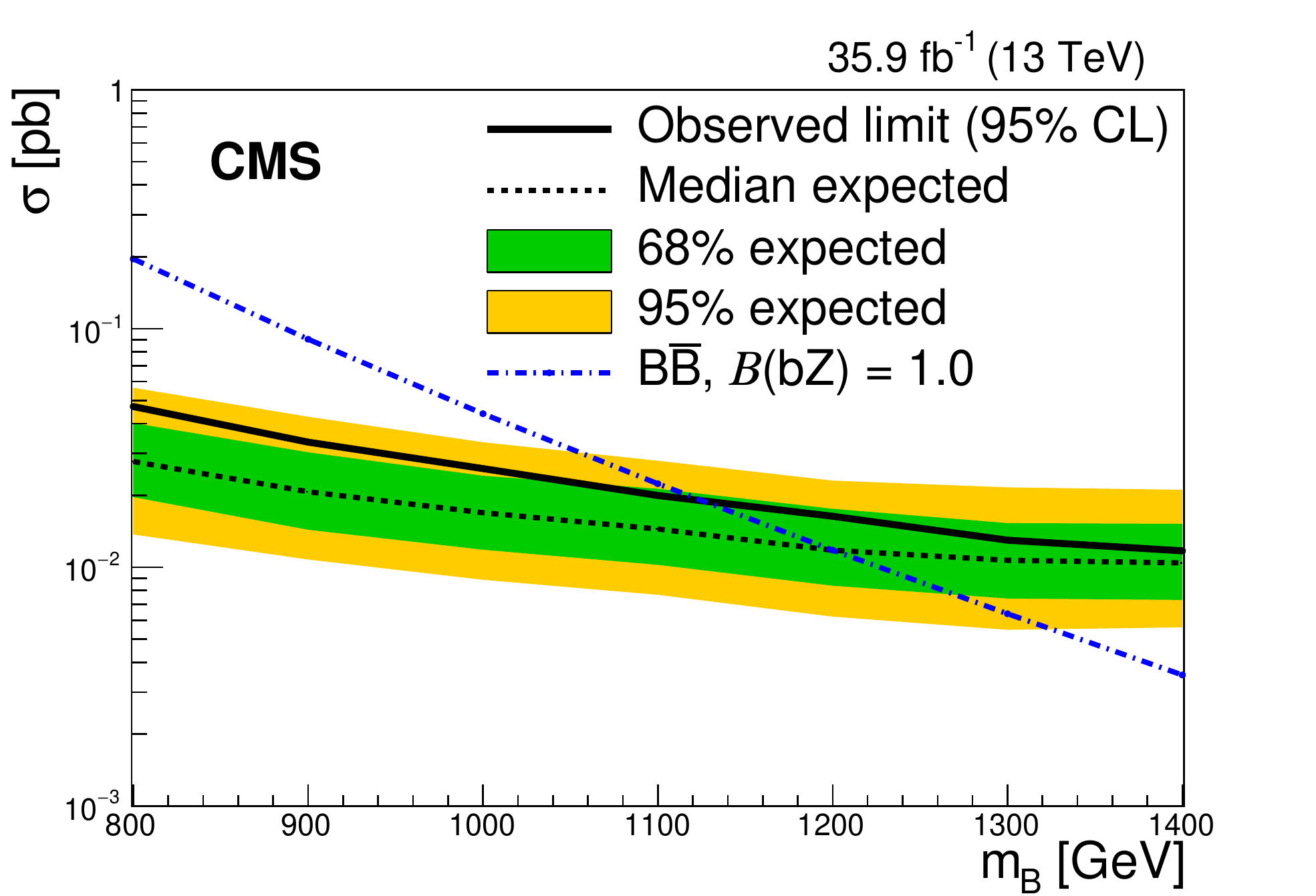}
    \includegraphics[width=0.48\textwidth]{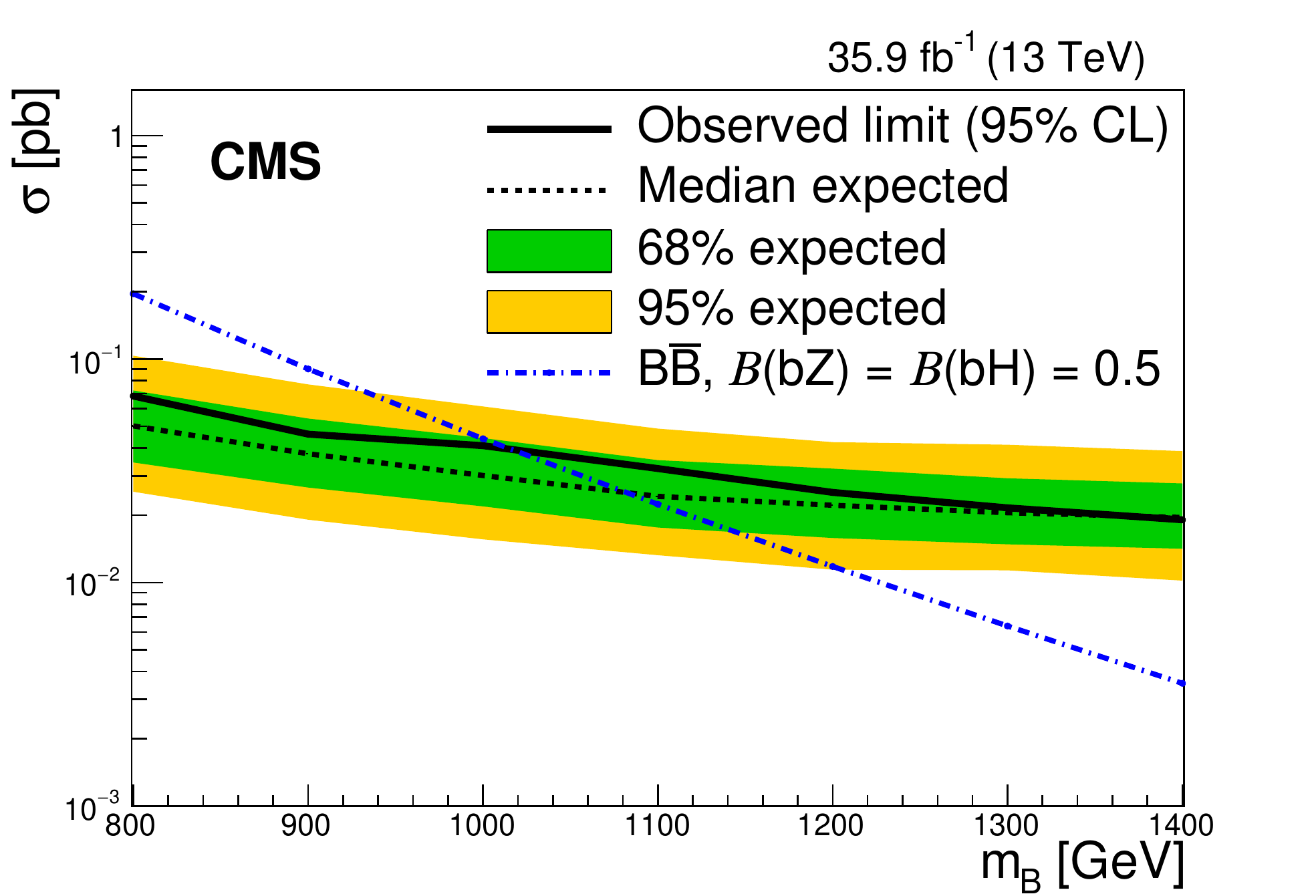}
    \includegraphics[width=0.48\textwidth]{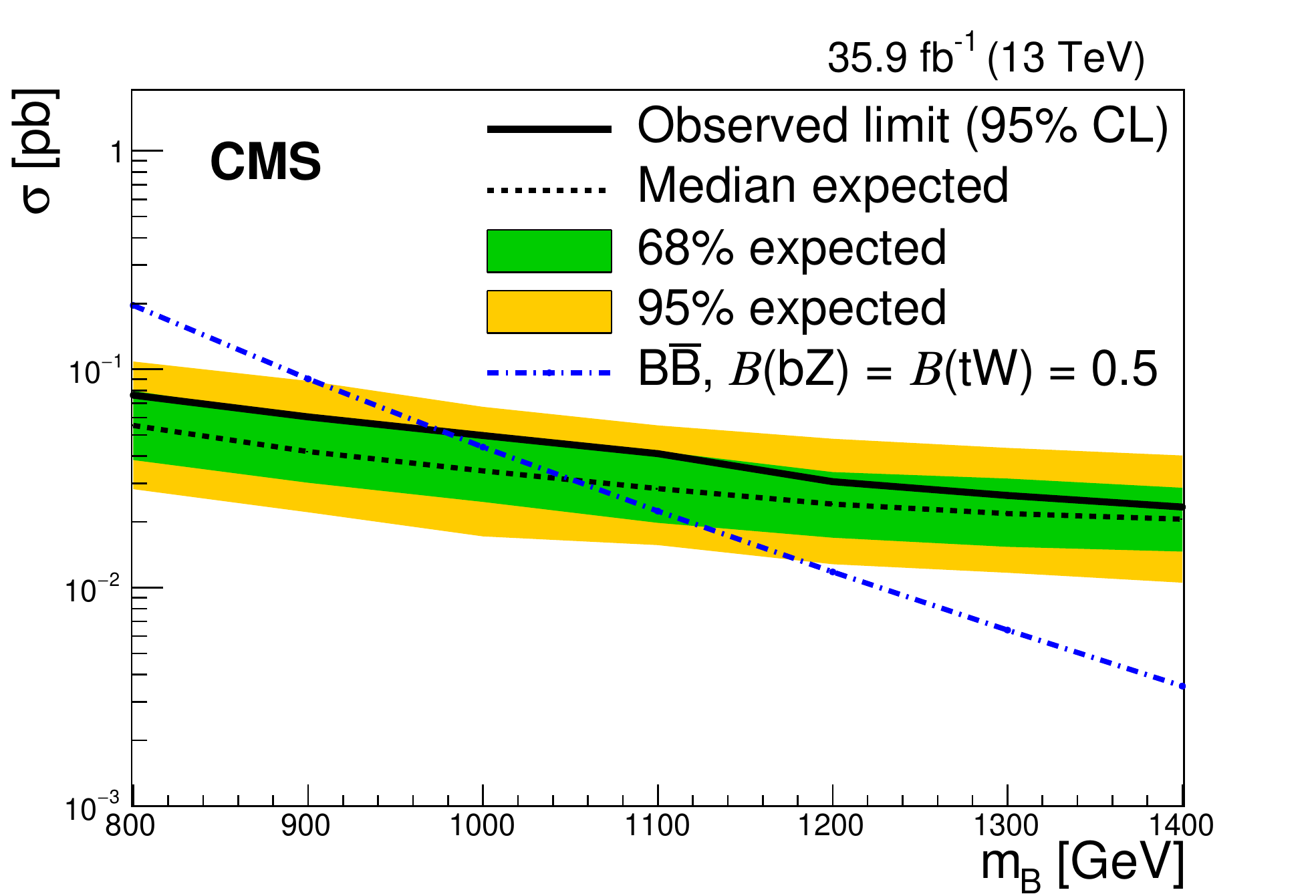}
     \caption{The observed (solid line) and expected (dashed line) 95\% \CL upper limits on the \BBbar production cross section versus the \B quark mass for (upper left) $\BR(\BtobZ) = 100\%$, (upper right) $\BR(\BtobZ) = \BR(\BtobH) = 50\%$, and (lower) $\BR(\BtobZ) = \BR(\BtotW) = 50\%$. The dotted-dashed line displays the theoretical cross section. The inner and outer bands show the one and two standard deviation uncertainties in the expected limits, respectively.}
    \label{fig:limit-B}

\end{figure*}

The upper limits at 95\% \CL on the \BBbar cross section are shown by the solid line in Fig.~\ref{fig:limit-B}. As before, the inner and outer bands give the one and two standard deviation uncertainties, respectively, in the expected upper limits. The dotted curve displays the theoretical signal cross section. Comparing the observed cross section limits to the signal cross section, we exclude \B quarks with masses less than 1130, 1015, and 975\GeV in the bZbZ, bZbH, and bZtW channels, respectively. The corresponding expected values are 1200, 1085, and 1055\GeV.

Figure~\ref{fig:limit-All} (lower) displays the observed (left) and expected (right) 95\% \CL lower limits on the \B quark mass as a function of the relevant branching fractions, assuming $\BR(\BtobZ) + \BR(\BtobH) + \BR(\BtotW) = 1.0$. For a \B quark decaying exclusively via \BtobZ, the lower mass limit is 1130\GeV.

\section{Summary}\label{sec:Summary}

The results of a search have been presented for the pair production of vector-like top (\T) and bottom (\B) quark partners in proton-proton collisions at $\sqrt{s} = 13\TeV$, using data collected by the CMS experiment at the CERN LHC, corresponding to an integrated luminosity of \intL~. The \TTbar search is performed by looking for events in which one \T quark decays via \TtotZ and the other decays via \TtobW, \tZ, \tH, where \PH refers to the Higgs boson. The \BBbar search looks for events in which one \B quark decays via \BtobZ and the other via \BtotW, \bZ, or \bH. Events with two oppositely charged electrons or muons, consistent with coming from the decay of a \PZ boson, and jets are investigated, and are categorized according to the numbers of top quark and \PW, \PZ, and Higgs boson candidates. These categories are individually optimized for \TTbar and \BBbar event topologies.

The data are in agreement with the standard model background predictions for all the event categories. Upper limits at 95\% confidence level on the \TTbar and \BBbar production cross sections are obtained from a simultaneous binned maximum-likelihood fit to the observed distributions for the different event categories, under the assumption of various \T and \B quark branching fractions. Comparing these upper limits to the theoretical predictions for the \TTbar and \BBbar cross sections as a function of the \T and \B quark masses, lower limits on the masses at 95\% confidence level are determined for different branching fraction scenarios. In the case of a \T quark decaying exclusively via \TtotZ, the lower mass limit is 1280\GeV, while for a \B quark decaying only via \BtobZ, it is 1130\GeV. These lower limits are comparable with those  measured by the ATLAS Collaboration~\cite{ATLAS-new}, also using the \PZ boson dilepton decay channel. The results of the analysis presented in this paper are complementary to previous CMS measurements~\cite{B2G-16-024,B2G-17-003,B2G-17-011}, and have extended sensitivity in reaching higher mass limits for \T and \B quarks.

\begin{acknowledgments}

We congratulate our colleagues in the CERN accelerator departments for the excellent performance of the LHC and thank the technical and administrative staffs at CERN and at other CMS institutes for their contributions to the success of the CMS effort. In addition, we gratefully acknowledge the computing centers and personnel of the Worldwide LHC Computing Grid for delivering so effectively the computing infrastructure essential to our analyses. Finally, we acknowledge the enduring support for the construction and operation of the LHC and the CMS detector provided by the following funding agencies: BMBWF and FWF (Austria); FNRS and FWO (Belgium); CNPq, CAPES, FAPERJ, FAPERGS, and FAPESP (Brazil); MES (Bulgaria); CERN; CAS, MoST, and NSFC (China); COLCIENCIAS (Colombia); MSES and CSF (Croatia); RPF (Cyprus); SENESCYT (Ecuador); MoER, ERC IUT, and ERDF (Estonia); Academy of Finland, MEC, and HIP (Finland); CEA and CNRS/IN2P3 (France); BMBF, DFG, and HGF (Germany); GSRT (Greece); NKFIA (Hungary); DAE and DST (India); IPM (Iran); SFI (Ireland); INFN (Italy); MSIP and NRF (Republic of Korea); MES (Latvia); LAS (Lithuania); MOE and UM (Malaysia); BUAP, CINVESTAV, CONACYT, LNS, SEP, and UASLP-FAI (Mexico); MOS (Montenegro); MBIE (New Zealand); PAEC (Pakistan); MSHE and NSC (Poland); FCT (Portugal); JINR (Dubna); MON, RosAtom, RAS, RFBR, and NRC KI (Russia); MESTD (Serbia); SEIDI, CPAN, PCTI, and FEDER (Spain); MOSTR (Sri Lanka); Swiss Funding Agencies (Switzerland); MST (Taipei); ThEPCenter, IPST, STAR, and NSTDA (Thailand); TUBITAK and TAEK (Turkey); NASU and SFFR (Ukraine); STFC (United Kingdom); DOE and NSF (USA).

\hyphenation{Rachada-pisek} Individuals have received support from the Marie-Curie program and the European Research Council and Horizon 2020 Grant, contract No. 675440 (European Union); the Leventis Foundation; the A. P. Sloan Foundation; the Alexander von Humboldt Foundation; the Belgian Federal Science Policy Office; the Fonds pour la Formation \`a la Recherche dans l'Industrie et dans l'Agriculture (FRIA-Belgium); the Agentschap voor Innovatie door Wetenschap en Technologie (IWT-Belgium); the F.R.S.-FNRS and FWO (Belgium) under the ``Excellence of Science - EOS" - be.h project n. 30820817; the Ministry of Education, Youth and Sports (MEYS) of the Czech Republic; the Lend\"ulet (``Momentum") Program and the J\'anos Bolyai Research Scholarship of the Hungarian Academy of Sciences, the New National Excellence Program \'UNKP, the NKFIA research grants 123842, 123959, 124845, 124850 and 125105 (Hungary); the Council of Science and Industrial Research, India; the HOMING PLUS program of the Foundation for Polish Science, cofinanced from European Union, Regional Development Fund, the Mobility Plus program of the Ministry of Science and Higher Education, the National Science Center (Poland), contracts Harmonia 2014/14/M/ST2/00428, Opus 2014/13/B/ST2/02543, 2014/15/B/ST2/03998, and 2015/19/B/ST2/02861, Sonata-bis 2012/07/E/ST2/01406; the National Priorities Research Program by Qatar National Research Fund; the Programa Estatal de Fomento de la Investigaci{\'o}n Cient{\'i}fica y T{\'e}cnica de Excelencia Mar\'{\i}a de Maeztu, grant MDM-2015-0509 and the Programa Severo Ochoa del Principado de Asturias; the Thalis and Aristeia programs cofinanced by EU-ESF and the Greek NSRF; the Rachadapisek Sompot Fund for Postdoctoral Fellowship, Chulalongkorn University and the Chulalongkorn Academic into Its 2nd Century Project Advancement Project (Thailand); the Welch Foundation, contract C-1845; and the Weston Havens Foundation (USA).
\end{acknowledgments}

\bibliography{auto_generated}

\cleardoublepage \appendix\section{The CMS Collaboration \label{app:collab}}\begin{sloppypar}\hyphenpenalty=5000\widowpenalty=500\clubpenalty=5000\vskip\cmsinstskip
\textbf{Yerevan Physics Institute, Yerevan, Armenia}\\*[0pt]
A.M.~Sirunyan, A.~Tumasyan
\vskip\cmsinstskip
\textbf{Institut f\"{u}r Hochenergiephysik, Wien, Austria}\\*[0pt]
W.~Adam, F.~Ambrogi, E.~Asilar, T.~Bergauer, J.~Brandstetter, M.~Dragicevic, J.~Er\"{o}, A.~Escalante~Del~Valle, M.~Flechl, R.~Fr\"{u}hwirth\cmsAuthorMark{1}, V.M.~Ghete, J.~Hrubec, M.~Jeitler\cmsAuthorMark{1}, N.~Krammer, I.~Kr\"{a}tschmer, D.~Liko, T.~Madlener, I.~Mikulec, N.~Rad, H.~Rohringer, J.~Schieck\cmsAuthorMark{1}, R.~Sch\"{o}fbeck, M.~Spanring, D.~Spitzbart, W.~Waltenberger, J.~Wittmann, C.-E.~Wulz\cmsAuthorMark{1}, M.~Zarucki
% ...
\vskip\cmsinstskip
% ...
1:  Also at Vienna University of Technology, Vienna, Austria\\
% ...
\end{sloppypar}
\end{document}